\documentclass[lettersize,journal]{IEEEtran}
\usepackage{amsmath,amsfonts}
\usepackage{amsthm}
\usepackage{bm}
\usepackage{algorithmic}
\usepackage{algorithm}
\usepackage{array}
\ifCLASSOPTIONcompsoc
    \usepackage[caption=false, font=normalsize, labelfont=sf, textfont=sf]{subfig}
\else
\usepackage[caption=false, font=footnotesize]{subfig}
\fi
\usepackage{textcomp}
\usepackage{stfloats}
\usepackage{url}
\usepackage{color, soul}
\usepackage{verbatim}
\usepackage{graphicx}
\usepackage{cite}
\usepackage{booktabs}
\usepackage{multirow}
\usepackage[flushleft]{threeparttable}
\usepackage[super]{nth}

\usepackage{soul}
\setstcolor{red}
\newcommand{\rev}[1]{\textcolor{black}{#1}} 
\definecolor{darkgreen}{rgb}{0.0, 0.5, 0.0} 

\begin{document}
\title{Electrifying Heavy-Duty Trucks: Battery-Swapping vs Fast Charging}
\author{Ruiting Wang, Antoine Martinez, Zaid Allybokus, Wente Zeng, Nicolas Obrecht, Scott Moura
\thanks{R. Wang and S. Moura. are with the Department of Civil and Environmental Engineering, University of California, Berkeley, CA, 94720 USA. \tt \{rtwang, smoura\}@berkeley.edu}
\thanks{A. Martinez is with École Polytechnique, Palaiseau, France \tt antoine.martinez.x21@polytechnique.edu}
\thanks{Z. Allybokus, W. Zeng, N. Obrecht are with TotalEnergies OneTech, Paris, France \tt\{zaid.allybokus, wente.zeng, nicolas.obrecht\}@totalenergies.com}
}

\markboth{IEEE Transactions on Smart Grid}%
{Shell \MakeLowercase{\textit{et al.}}: A Sample Article Using IEEEtran.cls for IEEE Journals}

\IEEEpubid{0000--0000/00\$00.00~\copyright~2021 IEEE}

\maketitle
\begin{abstract}
The advantages and disadvantages of Battery Swapping Stations (BSS) for heavy-duty trucks are poorly understood, relative to Fast Charging Stations (FCS) systems. This study evaluates these two charging mechanisms for electric heavy-duty trucks, aiming to compare the systems' efficiency and identify their optimal design. 
A model was developed to address the planning and operation of BSS in a charging network, considering in-station batteries as assets for various services. 
We assess performance metrics including transportation efficiency and battery utilization efficiency. Our evaluation reveals that BSS significantly increased transportation efficiency by reducing vehicle downtime compared to fast charging, but may require more batteries. BSS with medium-sized batteries offers improved transportation efficiency in terms of time and labor. FCS-reliant trucks require larger batteries to compensate for extended charging times. To understand the trade-off between these two metrics, a cost-benefit analysis was performed under different scenarios involving potential shifts in battery prices and labor costs.
Additionally, BSS shows potential for significant $\text{CO}_2$ emission reductions and increased profitability through energy arbitrage and grid ancillary services. 
These findings emphasize the importance of integrating BSS into future electric truck charging networks and adopting carbon-aware operational frameworks.
\end{abstract}

%


\newcommand{\softMax}{\text{sm}}
\newcommand{\logSumExp}{\text{lse}}

\newcommand{\LipschitzCst}{L}

\newcommand{\average}{\mathbb E}
\newcommand{\Proba}{\mathbb P}
\newcommand{\density}{p}

\newcommand{\Reals}{\mathbb R}
\newcommand{\Naturals}{\mathbb N}

\newcommand{\mult}{\odot}
\newcommand{\bydef}{\overset{\Delta}{=}}
\newcommand{\diff}{\Delta}
\newcommand{\subjectTo}{\text{s. to:} \quad }
\newcommand{\argmin}{\text{argmin}}
\newcommand{\argmax}{\text{argmax}}

\newcommand{\vecZero}{q}
\newcommand{\matZero}{Q}
\newcommand{\vecOne}{u}
\newcommand{\vecTwo}{v}

\newcommand{\iter}{k}
\newcommand{\Time}{t}
\newcommand{\stepSize}{\alpha}
\newcommand{\penaltyParam}{\rho}

\newcommand{\objective}{f}
\newcommand{\optimVar}{x}
\newcommand{\optimSet}{\mathcal X}
\newcommand{\optimVarSize}{n}
\newcommand{\constraintSize}{m}
\newcommand{\upperBound}{ub}
\newcommand{\lowerBound}{lb}
\newcommand{\optimum}{\star}
\newcommand{\problemParameters}{\theta}
\newcommand{\Aineq}{A}
\newcommand{\bineq}{b}
\newcommand{\Aeq}{C}
\newcommand{\beq}{d}

\newcommand{\activation}{\phi}

\newcommand{\sign}{\text{sgn}}
\newcommand{\constant}{C}
\theoremstyle{definition} 
\newtheorem{definition}{Definition}[section]
\newcommand{\scott}[1]{\hl{[SM: #1]}}


\section*{Nomenclature}
\subsection{Sets}
\begin{IEEEdescription}
\item[${\mathcal{I}}$] Set of station, battery combination with index $(s,i)$. Subset $\mathcal{I}_s$ represents batteries that are in station $s$. 
\item[$\mathcal{M}$] Set of customers with index $m$. 
\item[$\mathcal{M}_{s,t}$] Set of customers $m$, that visit station $s$ at time $t$.
\item[$\mathcal{O}$] Set of station, battery, customer combination with index $(s,i,m)$. 
\item[$\mathcal{S}$] Set of stations. Subset $\mathcal{S}_m$ represents the set of stations that customer $m$ visits. 
\item[$\mathcal{T}$] Set of time steps with index $t$. 
\item[$\mathcal{W}$] Set of station, customer combination with index $(s,m)$. 

\end{IEEEdescription}

\subsection{Integer Variables}
\begin{IEEEdescription}[\IEEEusemathlabelsep\IEEEsetlabelwidth{$x_{s,i,}$}]
\item[${x_{s,i,m}}$] Binary, if battery $i$ at station $s$ is selected and assigned for customer $m$. 
\item[${y_{s,i}}$] Binary, if battery $i$ at station $s$ is installed (for sizing of the stations). 
\item[$z_{s,i,t}$] Binary, if battery $i$ at station $s$ is swapped at time $t$. 
\item[${a_{i,t}}$] Binary, indicator of charging or discharging. 
\item[${b_m}$] Binary, indicator of whether a customer received a battery swap. 
\end{IEEEdescription}

\subsection{Continuous Variables}
\begin{IEEEdescription}[\IEEEusemathlabelsep\IEEEsetlabelwidth{${q^{\text{s2v}}_{m,}}$}]
\item[${p^{\text{chg}}_{s,i,t}}$] Charging rate of battery $i$ at station $s$ at time $t$. 
\item[${p^{\text{dsg}}_{s,i,t}}$] Discharging rate of battery $i$ at station $s$ at time $t$. 
\item[${e^{\text{G}}_t}$] Energy bought from the grid at time $t$. 
\item[${q_{s,i,t}}$] Energy level of battery $i$ at station $s$ at time $t$. 
\item[${q^{\text{s2v}}_{m,s}}$] Energy sold to customer $m$ at station $s$. 
\item[${q^{\text{veh}}_{m,s}}$] Energy level of the Vehicle-to-Station (V2S) battery from customer $m$ when it arrives at station $s$.
\item[${q^{\text{shrt}}_{m,s}}$] Energy shortage of the Station-to-Vehicle (S2V) battery provided to customer $m$ at station $s$. 
\end{IEEEdescription}
\IEEEpubidadjcol

\subsection{Parameters}
\begin{IEEEdescription}[\IEEEusemathlabelsep\IEEEsetlabelwidth{${\text{SOE}^{\text{ex}}_{m}}$}]
\item[$\alpha_1, \alpha_2$] The threshold of the minimum percentage of customers that must receive a swap, and energy demand that must be satisfied. 
\item[$B^{\text{cap}}$] Capital investment of the battery averaged over the entire lifetime to one day. 
\item[$t_b, t_e$] The first and the last time step. 
\item[${F^{\text{G}}_t}$] Electricity price at time $t$. 
\item[${R^{\text{sw}}}$] Revenue from per swap. 
\item[${R^{\text{en}}_t}$] Per kWh energy revenue at time $t$. 
\item[$P^{\text{en}}$] Penalty from energy shortage. 
\item[$\eta$] Charging/discharging efficiency. 
\item[$\overline{P}$] Maximum charging/discharging power. 
\item[$\overline{E}$] Grid energy exchange limits in one time step. 
\item[$Q$] Battery capacity in kWh. 
\item[$Q^{\text{ini}}_{m}$] Initial energy level of the battery in the vehicle of customer $m$. 
\item[$C_{m,s}$] Energy consumption in kWh of customer $m$ from origin to station $s$.
\item[${E_{m,s}}$] Energy required by customer $m$ at station $s$. This value measures the energy required for each truck to complete the assigned route. 
\item[$T_{m,s}$] Arrival time of customer $m$ at station $s$. 
\item[$N$] Total number of customers/trips. 
\rev{\item[$\xi_t^{\text{CO}_2}$] $\text{CO}_2$ emission factor at time $t$. This value could be average emission factor $\xi_t^{\text{Avg}}$ or marginal emission factor $\xi_t^{\text{Mar}}$}
\end{IEEEdescription}

\section{Introduction}\label{sec:intro}

\subsection{Background}

Electrifying the trucking fleet has the potential to substantially reduce emissions in the transportation sector.
In the U.S. Environmental Protection Agency (EPA) rule published in March 2024, ``Greenhouse Gas Emissions Standards for Heavy-Duty Vehicles – Phase 3,'' the standards aim to reduce greenhouse gas (GHG) emissions from heavy-duty vehicles by 29\% below 2021 levels by 2032. This standard is by far the strongest and most stringent of its kind, pushing the industry to take actions in electrification to meet the new standards \cite{EPA2023, EPA2024}.  

This is, however, a challenging task. 
Weight limits of heavy-duty trucks (HDTs) mean that operators must trade off payload against battery size due to lithium-ion cells' limited gravimetric energy density. In addition, the driving range and the vehicles' capability to satisfy customer demands are strongly limited by the long recharge time. While rapid charging technologies have improved significantly over the past decade, battery charging still takes significantly more time than traditional diesel refueling \cite{liimatainen2019}. As a result,  \rev{this extended downtime can impact operational efficiency in time-sensitive logistics operations, potentially requiring} more than one electric truck to replace a diesel truck \rev{to maintain the same service levels} \cite{wang2024b, cabukoglu2018}. 
Meanwhile, fuel cell trucks have a relatively low ``well-to-wheels'' energy conversion efficiency compared to electric trucks, leading to possibly worse GHG emissions per mile compared to diesel. Although the feasibility of electric trucks is greatly improved by the proliferation of fast charging infrastructures, concerns include high upfront costs, long queues during peak occupancy, adverse impacts on battery health, and challenges on the power grid stability \cite{pelletier2017,tomaszewska2019}. 

Some companies, such as Heliox \cite{heliox}, Kempower \cite{Kempower}, and Volvo \cite{volvo} propose building direct current (DC) fast charging networks to support electric truck fleets for quick turnaround times. In Sweden, for example, there are plans to open a total of 130 DC fast charging stations (FCS) in 2023 and 2024 \cite{volvo}. In 2023, the Biden-Harris Administration in the U.S. announced a \$400 million commitment to deploy over 1,000 DC fast-chargers for heavy-duty electric trucks along freight corridors in California in the next decades \cite{biden}. 

Battery-swapping, as an alternative, has attracted attention because of the fast swap time and other potential benefits, such as allowing trucks to be sold without batteries, reducing peak demand by temporally decoupling charging and mobility, and introducing economic benefits by providing grid services, etc. While standardization remains a key challenge requiring collaborative effort from multiple stakeholders, recent studies have demonstrated the economic feasibility of BSS \cite{li2024, chen2022}. However, as will be shown in the following section, the optimal charging mechanism for long-haul HDTs in the context of logistics industry operations is not well understood.

\subsection{Literature Review}
\label{sec:lr}

In operation, the main difference between battery swapping and other charging mechanisms such as fast charging is that battery swapping decouples the temporal relationship between customer en-route charging demand and the demand to the power grid. It also offers a potential solution to overcome the operational drawbacks of electric trucks, with a service time comparable to or even beating traditional diesel trucks \cite{arora2023}. This eliminates the need to make significant changes to original routes due to the increased charging time requirements and tight customer service windows in the electrification of the business. 

The concept of BSS has experienced many setbacks and failures in the past, from Betterplace's bankruptcy in 2013, to Tesla abandoning their battery-swap program \cite{chen2022}. However, there are also some successful deployments. In China, the market share of battery-swapping electric trucks surged from 0.24\% to 1.9\% in 2022, and accounts for 51.3\% of the electricity-based trucks that year \cite{china2023}. Article \cite{chen2022} provides an overview of the concepts, architectures, implementations, and standardization issues of BSS.

Many recent studies focus on addressing challenges in commercializing BSS and analyzing the business cases for HDTs. Reference \cite{zhang2024a} compared fast charging and battery swapping charging services, using slow charging as a benchmark, in terms of pricing and profitability. Articles \cite{li2024, zhu2023} perform a technological-economic analysis of the operation of BSS for HDTs based on data available in battery-swapping pilot cities, and demonstrate cost-effective battery swapping modes. Other works analyze the operating cost \cite{wu2021a}, and perform a simplified life cycle assessment \cite{yang2018a} with different settings. In these studies, individual stations are modeled in detail. However, studying the interactions within a broader BSS network for freight applications has received less attention.

Studies related to BSS system design and infrastructures are mainly focusing on algorithm design and are indifferent to the vehicle types. There are two main types of BSS systems: centralized BSS, where charging and swapping services may not occur at the same location, and distributed BSS, where both services occur at the same location. 
A centralized BSS system decouples transportation demand from power grid demand in terms of time and location, by using a battery logistics network to transport batteries from central charging stations to swapping branches. This results in a more complex system that requires a separate truck fleet solely for transporting batteries and necessitates daily operations. The research reported in \cite{liu2019, ban2019a, tan2019, zhu2024} has proposed different algorithms for the optimal scheduling of such a system. The article by Qi \emph{et al.} addresses the infrastructural challenges, and proposes a joint location and repairable inventory model \cite{qi2023}. Zhu \emph{et al.} study the decarbonization of the battery logistics system as part of the multiple integrated energy systems \cite{zhu2024}.
For distributed BSS, an early paper in infrastructure design is \cite{mak2013}, where robust optimization models were proposed to aid the planning process of BSS. Other papers study the operating model with battery-to-grid features \cite{sarker2015}; optimal sizing of such a system \cite{siddiqua2023}; BSS's potential for fast frequency regulation services \cite{wang2021}; BSS for electric buses in distribution systems \cite{zheng2014, el-taweel2023}, etc. Some papers also discuss the BSS operation problem coupled with routing \cite{yang2015, raeesi2020, hof2017, liang2023, mao2024}, with privacy-preserving requirement \cite{wan2023}, etc. 

Our work is developed in a distributed BSS network scenario for HDTs. This networked perspective is particularly crucial for heavy-duty freight applications where long-haul routes require coordinated planning across geographically dispersed stations. While battery swapping operation and scheduling modeling methods and algorithms have been studied for a decade for passenger cars, there has not been dedicated research focusing on BSS for HDTs in logistic networks. 
\rev{The unique challenges of BSS networks for freight transportation--including different arrival patterns, diverse travel behavior, and the interdependencies between stations along major freight corridors--introduce complexities that existing models don’t fully address.} 
No study has comprehensively examined the system efficiency of heavy-duty electric trucks by comparing various charging mechanisms in a regional charging network--a main focus of this work.


\subsection{Statement of Contribution}

This work expands on our previous conference paper \cite{rtw_ACC} by examining BSS operation in a geographic network. We address critical issues such as station sizing, optimal battery capacity, and system efficiency for both BSS and FCS using real demand data from California, USA. Our goal is to understand and compare the characteristics of electric long-haul heavy-duty trucks under different charging mechanisms. Our key contributions are as follows:

\begin{enumerate}
    \item \rev{Network optimization}: We propose innovative optimization models for the strategic planning of infrastructure, sizing of stations, and efficient daily operation of BSS systems \rev{across regional freight corridors, addressing the unique demands of heavy-duty logistics networks.}
    \item Efficiency evaluation: We \rev{quantify and }compare the transportation efficiency and battery utilization efficiency \rev{between BSS and FCS technologies across multiple battery configurations, revealing critical performance differentials for heavy-duty applications.}
    \item Multi-Services: \rev{We assess the additional value streams from BSS implementations, including $\text{CO}_2$ emission reduction potential and economic benefits from grid services, establishing a comprehensive framework for evaluating charging infrastructure investments.}
\end{enumerate}

\subsection{Outline}
Section \ref{sec:formulation} introduces the problem formulation we used for this work; Section \ref{sec:exp} provides an overview of the dataset and parameters setting of numerical experiments; Section \ref{sec:result} presents our findings on the best battery sizing, efficiency evaluations, and operation analysis of BSS; and finally, Section \ref{sec:conclu} discusses the potential insights and concludes the study. 
\section{Problem Formulation}\label{sec:formulation}

\subsection{Assumptions}

In this study, we assume that a single entity operates multiple BSSs within the network. This setup allows for information exchange and coordination of swapping services. The day-ahead assignment of truck routes provides complete knowledge of relevant routing details, including energy demand and expected arrival times.  In addition, it is assumed that swapping occurs in a single time step. This study primarily uses a time step of 0.5 hours. Based on existing industry battery swapping applications, each swap typically takes 5–10 minutes \cite{li2024, ample_bss}. The remaining time in each time step acts as a buffer, accounting for the total duration that trucks spend off-highway. 

\subsection{Objectives}

The present study considers two stages of the problem. The stages include the initial system design to the daily operational phase, where different objectives are employed at each stage. 

\subsubsection{Planning Problem -- Minimize battery number}

In order to determine the optimal battery configuration for the entire network, we minimize the total number of batteries required, subject to feasibility constraints that will be introduced later. The decision variable $y_{s,i}$ is used to indicate whether battery $i$ in station $s$ is installed. $y_{s,i}$ is $1$ only if battery $s, i$ has been active in any battery swapping events.

\begin{align}\label{obj:min_bat}
   \min \sum_{s,i \in \mathcal{I}} y_{s,i}
\end{align}


\subsubsection{Operations Problem -- Minimize operational cost}

For daily operations, the following prices are considered as known parameters: $F^{\text{G}}_t$ is per kWh price at time $t$, when we buy and sell energy from and to the grid. Parameter $R^{\text{sw}}$ is the revenue per swap, and $R^{\text{en}}_t$ is the per kWh revenue when providing a swapping service to a customer with the demand of $E_{m,s}$ at time $t$. When we violate the customer's request due to the inability to provide a battery with sufficient energy for swapping, a penalty would occur to represent the value of customer dissatisfaction. For each kWh of energy shortage, there is a penalty of $P^{\text{en}}$. During operation, we seek to minimize the following objective function:

\begin{align}\label{obj_operation}
\begin{split}
\min \quad & \rev{J_\text{op}}, \quad\text{where}\\
& \rev{J_\text{op} = \sum_{t\in \mathcal{T}} 
F^{\text{G}}_t  e^{\text{G}}_t 
-R^{\text{sw}} \sum_{s,i \in \mathcal{I}, t \in \mathcal{T}} z_{s,i,t}} \\ 
&\rev{- \sum_{s, m \in \mathcal{W}} R^{\text{en}}_{T_{m,s}} q^{\text{s2v}}_{m,s} 
+ \sum_{s, m \in \mathcal{W}}
P^{\text{en}} q^{\text{shrt}}_{m,s} }
\end{split}
\end{align}
The remaining notation is defined in the Nomenclature section.

\subsection{Constraints}

\subsubsection{Swapping Choice}
Binary variable $x_{s,i,m}$ is the key decision variable that captures the occurrence of a swapping service and the assignment of batteries between battery $(s,i)$ and customer $m$. When the value of $x_{s,i,m}$ is 1, battery $i$ at station $s$ is selected and assigned for customer $m$. 

Binary variable $z_{s,i,t}$ shows whether the battery $(s,i)$ is swapped at time $t$. The relationship between these two variables can be captured by the following constraints \eqref{sig_bat_demand}. Each battery can only be assigned to a single customer at a certain time step. 

\begin{equation}\label{sig_bat_demand}
    \sum_{m\in \mathcal{M}_{s,t}} x_{s,i,m} = z_{s,i,t} \quad \forall s, i\in \mathcal{I}, \forall t \in \mathcal{T},
\end{equation}

It is assumed that each customer will only swap once when visiting a station, described by constraint \eqref{sig_cus_demand}. 

\begin{equation}\label{sig_cus_demand}
    \sum_{i \in \mathcal{I}_s} x_{s,i,m} \leq 1 \quad 
    \forall s,m \in \mathcal{W}
\end{equation}

It is not permitted to perform two consecutive swaps for the same battery within two adjacent time step intervals. To clarify, the ``vehicle-to-station'' \rev{(V2S)} battery that has been swapped into the station will be charged for at least one time step before it can be swapped to another customer. This is expressed by constraint \eqref{no_two_subsequent_swap}. 

\begin{equation}\label{no_two_subsequent_swap}
    z_{s,i,t} + z_{s,i,t+1} \leq 1 \quad \forall s, i\in \mathcal{I}, \forall t \in \mathcal{T}/\{t_e\},
\end{equation}


\subsubsection{State-of-Energy Dynamics}
The state-of-energy of the battery is updated based on the charging/ discharging power during the hour if not swapped. Otherwise, it would be updated by the energy of the V2S battery, with a value of $q^{\text{veh}}_{m,s}$, shown in constraint \eqref{update_soc}. Similarly, we update the actual energy sold to customers $q^{\text{s2v}}_{m,s}$ based on battery assignment in constraint \eqref{update_soc_sell}. The big-M method is used for these constraints for linearization. In this case, a $M$ equal to battery \rev{nominal} capacity $Q$, is sufficient. 

\begin{align}\label{update_soc}
\begin{split}
q_{s,i,t+1} = 
\begin{cases} 
      \sum_{m\in \mathcal{M}_{s,t}}{q^{\text{veh}}_{m, s} x_{s,i,m}} & z_{s,i,t} = 1\\
      q_{s,i,t} +  \Delta t \left( p^{\text{chg}}_{s,i,t} \eta - p^{\text{dsg}}_{s,i,t}/{\eta} \right) & o.w. \\
   \end{cases} 
   \\ \forall s, i\in \mathcal{I}, \forall t \in \mathcal{T},
\end{split}
\end{align}

\begin{equation}\label{update_soc_sell}
\begin{split}
q^{\text{s2v}}_{m,s} = 
\begin{cases} 
      q_{s,i,T_{m,s}} - q^{\text{veh}}_{m, s} & x_{s,i,m} = 1 \\
      0 & o.w. \\
   \end{cases} 
   \\ \forall s, i, m \in \mathcal{O},
\end{split}
\end{equation}

\rev{We assume a constant charging efficiency for simplicity and to preserve the linearity of the optimization problem. However, actual charging efficiency is inherently nonlinear, as it varies with the state of charge (SoC), temperature, and charging power level \cite{saxena2015}. Nonlinear efficiency effects, particularly at high SoC and during fast charging, arise due to increased internal resistance and heat losses \cite{ansean2013}.} 

\subsubsection{Customer Demand}

In constraint \eqref{SOE_demand}, the energy shortage of the ``station-to-vehicle'' (S2V) batteries is measured,  

\begin{equation}\label{SOE_demand}
\begin{split}
    q^{\text{shrt}}_{m,s} \geq E_{m,s} x_{s,i,m} - q_{s,i,T_{m,s}} 
    \\\quad \forall s, i, m \in \mathcal{O},
\end{split}
\end{equation}

If no exchange service is provided along the entire route, then the energy shortage is measured by the maximum energy storage at each station by comparing the energy level of the vehicle at station $s$ and its energy demand, as in constraint \eqref{SOE_no_swap},
\begin{equation}\label{SOE_no_swap}
\begin{split}
    q^{\text{shrt}}_{m,s} \ge  E_{m,s} \left(1 - \sum_{s,i \in \mathcal{I}} x_{s,i,m} \right) - q^{\text{veh}}_{m,s}
    \\\quad \forall s, m \in \mathcal{W},
\end{split}
\end{equation}

In the constraints \eqref{SOE_vehicle}, the energy level of the vehicle of customer $m$ in station $s$, $q^{\text{veh}}_{m,s}$ is determined by the energy consumption of that route as well as the amount of energy recharge from the swapping in the previous (if any) station $s'$. The set $\mathcal{W}_k$ represents the station and customer combination that indicates the first stop with no previous swap. The set $\mathcal{W}_h$, similarly, is the set of second stops. 

\begin{align}\label{SOE_vehicle}
\begin{split}
q^{\text{veh}}_{m,s} = 
\begin{cases} 
      Q^{\text{ini}}_{m} - E_{m,s} & \forall s, m \in \mathcal{W}_k\\
      Q^{\text{ini}}_{m} - E_{m,s} + q^{\text{s2v}}_{m,s'}  & \forall s, m \in \mathcal{W}_h\\
   \end{cases} 
\end{split}
\end{align}

\subsubsection{Energy Equality}

Constraint \eqref{energy_balance} shows that the total energy bought from and sold to the grid is determined by aggregating all the charge and discharge energy of the battery. 

\begin{equation}\label{energy_balance}
    \sum_{s,i \in \mathcal{I}} \left(p^{\text{chg}}_{s,i,t} - p^{\text{dsg}}_{s,i,t} \right) \Delta t= e^{\text{G}}_t \quad \forall t \in \mathcal{T},
\end{equation}


Constraints \eqref{char_dischar1} and \eqref{char_dischar2} ensure that charging and discharging of a single battery do not occur simultaneously.
Constraints \eqref{char_dischar1}-\eqref{grid} also set the maximum charging/discharging powers, and grid energy exchange upper bound \rev{is} limited by the \rev{maximum power ratings} of the chargers and converters.
 
\begin{equation}\label{char_dischar1}
    p^{\text{chg}}_{s,i,t} \le \overline{P} a_{i,t} \quad \forall s, i\in \mathcal{I}, \forall t \in \mathcal{T},
\end{equation}
\begin{equation}\label{char_dischar2}
    p^{\text{dsg}}_{s,i,t} \le \overline{P} (1-a_{i,t}) \quad \forall s, i\in \mathcal{I}, \forall t \in \mathcal{T},
\end{equation}
\begin{equation}\label{grid}
    |e^{\text{G}}_t| \le \overline{E}  \quad \forall t \in \mathcal{T},
\end{equation}

\subsubsection{Battery Sizing}

For station sizing, we introduce the variable $y_{s,i}$ and propose the following constraints.
Constraint \eqref{sizing1} sets the indicator variable $y_{s,i}$ to $1$ only if battery $s, i$ has been active with battery swapping events, that is, the battery is needed for the system.
Constraint \eqref{sizing2} makes sure that if it is not active, it stays at zero energy level. 

\begin{equation}\label{sizing1}
     \frac{1}{M} \sum_{m\in \mathcal{M}} x_{s,i,m} \le y_{s,i}  \le \sum_{m\in \mathcal{M}} x_{s,i,m} \quad \forall s, i \in \mathcal{I},
\end{equation}
\begin{equation}\label{sizing2}
     q_{s,i,t}  \le  Q y_{s,i}  \quad \forall s,i \in \mathcal{I},  \forall t \in \mathcal{T}
\end{equation}

\subsubsection{Customer Satisfaction}

Two distinct sets of constraints can be employed to ensure the occurrence of specific levels of customer dissatisfaction, by \eqref{100_served}, or by \eqref{cus_m}-\eqref{pct_energy_satis}. All these constraints are optional and may be disregarded if customer satisfaction rate is not a concern in operation.

To guarantee 100\% of the customers are served, we use constraint \eqref{100_served}, 

\begin{align}\label{100_served}
    q^{\text{shrt}}_{m,s} = 0 \quad \forall s, m \in \mathcal{W}
\end{align}

It is also possible to configure two thresholds to restrict the minimum percentage of customers that must receive a swap (constraint \eqref{num_cus_satis}) and the minimum percentage of energy demand that must be satisfied (constraint \eqref{pct_energy_satis}). The constraint \eqref{cus_m} is employed to ascertain whether customer $m$ is provided with a battery. 

\begin{align}\label{cus_m}
    \sum_{s\in S_m} q^{\text{shrt}}_{m,s} \leq M b_m \quad \forall m \in \mathcal{M}, 
\end{align}
\begin{align}\label{num_cus_satis}
    \sum_{m \in \mathcal{M}} b_m \geq \alpha_2 N \quad \forall  s, m \in \mathcal{W}
\end{align}
\begin{align}\label{pct_energy_satis}
    q^{\text{shrt}}_{m,s} \leq (1-\alpha_1) E_{m,s} \quad \forall  s, m \in \mathcal{W}
\end{align}

\subsection{Formulation Summary}
To summarize, in the planning problem, we minimize \eqref{obj:min_bat}, s.t. \eqref{sig_bat_demand} - \eqref{100_served}. 

In the operation problem, we minimize \eqref{obj_operation}, s.t. \eqref{sig_bat_demand} - \eqref{grid}, and \eqref{cus_m} - \eqref{pct_energy_satis}.

\section{Numerical Experiment Setup}\label{sec:exp}
\subsection{Dataset}

In this section, we first provide an overview of the truck trip dataset used for this study. We then introduce our energy estimation methods, and present the truck trip characteristics. Finally, the section provides the necessary parameters for trucks and stations.

\subsubsection{Trip Dataset}
The NextGen NHTS dataset has been employed for the generation of use cases \cite{NHTS}. This paper leveraged the annual passive original destination (OD) truck data products and focused on truck trips exceeding 300 miles in length, within the state of California. California is divided into 26 Metropolitan Statistical Areas (MSA) and 4 Micropolitan Statistical Areas (MiSA, or NonMSA). These areas are defined and updated by the Office of Management and Budget (OMB) on February 28, 2013 based on the application of the new standards to data from the 2010 Census \cite{MSA_area}. These areas are determined based on population and degree of social and economic integration in adjacent territories. We examine internal daily truck trips within these zones. 

\subsubsection{Trip Metrics Estimation}

Among California's 30 NHTS geographical zones, we query trip distance, travel time, and route elevation using the HERE Truck Routing API \cite{HEREDeveloper}. The longitudinal dynamics of vehicles have been considered for energy prediction. We assume constant acceleration and deceleration at the beginning and ending stages of the speed profile, respectively, and constant speed in between. We utilized the energy estimation model proposed by Basso et al. \cite{basso2019}, which is also summarized in Appendix \ref{energy_model}.

\subsubsection{Truck and Station Specifications}

It is assumed that the specific energy of the battery is 680 kg ($\sim$ 1500 lb)/100 kWh, \rev{or, 147 Wh/kg, consistent with }reference \cite{weight2022a}. \rev{We tested a range of battery sizes from 350 kWh to 800 kWh.}
For the vehicle, we assume the curb weight without battery is 2268 kg (5000 lb) \cite{weight2022a}. Additionally, the GVWR limitation for electrified heavy-duty trucks is 82,000 lb by the U.S. Department of Transportation. It is assumed that all trucks are fully loaded at all times and reach the GVWR limit. This is not true, in practice. Consequently, the following results correspond to a worst-case scenario. Drivers are required to comply with the intrastate hours-of-service rules with a maximum duty period of 12 hours, and must be at least 10 hours off-duty afterwards\footnote{California does not have a 30-minute rest break requirement for intra-state trips.}\cite{ca_hos}. 
Drivers may drive for up to 12 hours per day, which represents the maximum number of hours that a driver may work in a single day \cite{driving_hour}. For fast charging, a 1 C charging rate and a maximum charging power of 750 kW are assumed, given the existing charger capacity for private cars in the market \cite{shi2022} and similar studies on electric trucks \cite{speth2022}. For battery swapping, we take a 1/3 C charging rate for batteries in stations. In Appendix \ref{char_speed}, other charging speeds are evaluated. 

\rev{In the analysis, we consider FCS and BSS primarily in the context of large-scale infrastructure planning to meet trucking needs, assuming them as stand-alone stations exclusively for heavy-duty trucks that fulfill the same amount of customer demands. This approach allows for direct comparison of their operational characteristics, though we recognize that real-world implementations might benefit from shared infrastructure models that could significantly alter the economic equation.} 

\subsection{BSS Penetration and Rerouting}

\begin{figure}[htbp]
\centering
\subfloat[The distance/energy consumption of original trips that require charging, of feasible trips with a maximum of four stops, and of routed trips after considering recharge detour for different battery sizes.]
{\includegraphics[width = \linewidth]{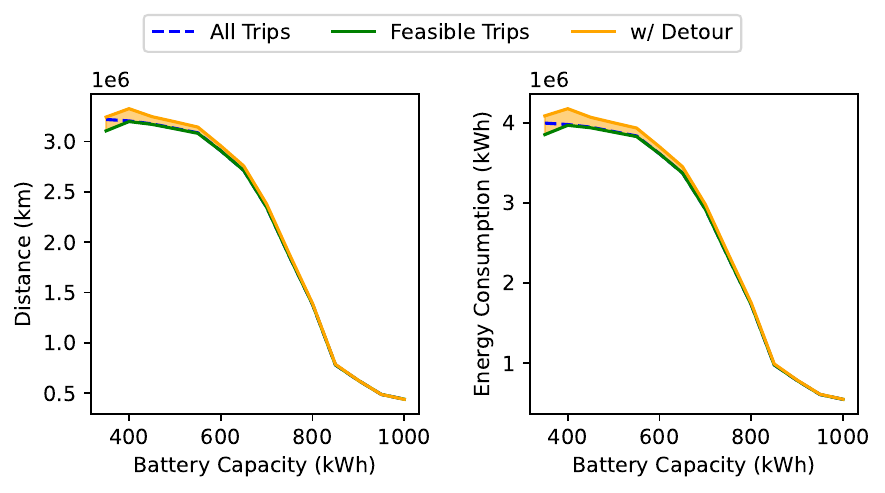}} \\
\subfloat[Daily truck trip energy consumption distribution in California. ] 
{\includegraphics[width = \linewidth]{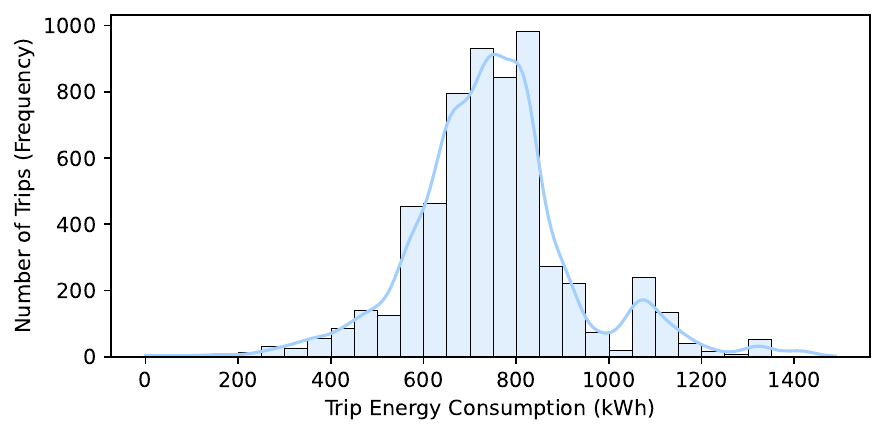}}
\caption{Daily truck trip energy distance and energy consumption under different conditions and the energy demand considered in the experiments. }
\label{fig:dist_energy}
\end{figure}

In the context of electrifying truck fleets, we consider battery swapping exclusively for en-route charging. Namely, we propose a combination of BSS for en-route recharging and slow charging for overnight depot charging. This suggests that truck fleets with larger battery sizes will have reduced needs for recharging en-route. This is illustrated in Fig. \ref{fig:dist_energy}(a). As the battery size on the truck increases, the total distance and the total energy consumption of all trips (blue dotted line) that require en-route recharging or battery swapping decreases. 

The trips that do not require recharging en-route are grouped together as round trips. For example, if there are multiple requests for trip A-B, the trips are processed as A-B-A-B-A-B... until all requests are satisfied or the energy consumption exceeds the battery capacity, in which case another truck would be dispatched. 

The daily truck trip energy consumption distribution in California is shown in Fig. \ref{fig:dist_energy}(b). En-route recharging and battery swapping only happen when the battery size is less than the total energy consumption required for daily trips. 

When en-route battery swapping is required for a particular trip, trucks need to re-route to a station for battery swapping. Given the specified granularity and the imposed limit on the number of stops, it is possible that the truck may not always be able to identify a feasible trip. A quantitative analysis was conducted to determine the percentage of intra-zone truck trips that can be satisfied by a given maximum number of stops, with varying battery sizes. In this context, the term ``BSS penetration rate'' will refer to this constraint. 
In the case of four-stop routes, it is possible to satisfy more than 97.0\% of trips with a battery capacity of 350 kWh, and more than 99.7\% of demand with a battery capacity larger than 350 kWh. When the battery capacity is at least 600 kWh, 100\% of the trips can be fulfilled. Battery swapping is thus permitted a maximum of four times per route. 


In numerical experiments, we permit a maximum of four stops for routing, which results in the distance and energy consumption shown in Fig. \ref{fig:dist_energy}. The original trips include all trips that require recharging with the corresponding battery capacity. The feasible trips are those trips that can be successfully electrified for BSS and rerouted for charging. We record the total distance of these trips with and without the charging detour. The orange-shaded region represents the increased distance and energy consumption due to the detour for charging.

\subsection{Electricity Market}
Besides providing electric mobility services, we also consider providing grid services from a BSS. The grid-side price signal is obtained from the hourly Locational Marginal Prices (LMP) in the day-ahead market (DAM) from the California Independent System Operator (CAISO) \cite{caiso} Open Access Same-time Information System (OASIS) \cite{oasis}. 

The $\text{CO}_2$ emissions signals are obtained from WattTime \cite{watttime}. The first weeks of March, June, September, and December in 2023 are used to study seasonal variation. Several baselines are considered in setting the $\text{CO}_2$ cost range. We adopted the definition in 
\cite{nordhaus2017, auffhammer2018}, where they evaluate the social cost of carbon (SCC) as the marginal social damage from emitting one metric ton of $\text{CO}_2$-equivalent at a certain time. Bressler \cite{bressler2021} estimates the number of deaths caused by the emissions of one additional metric ton of $\text{CO}_2$, and converts them to economic values, referred to as the mortality cost of carbon (MCC). In the baseline emission scenario reported in the year 2020, SCC is \$37 and MCC is \$258 \cite{bressler2021}. These values exhibit variability depending on the specific scenarios. The presented analysis aims to encompass the full range of potential values, considering a carbon cost from 0 to 1,000 USD per ton. 

When not serving trucking customers, we use the BSS to provide the grid ancillary services (AS). The present study considers frequency regulation services, and incorporates regulation down (RD) and regulation up (RU) in the model. AS clearing prices for DAM are collected as price signals for the provision of grid services. A series of numerical experiments was conducted in Section \ref{sec:as} for the day-ahead market in the first week of March, June, September, and December in the year 2023. The results of the numerical experiments conducted are presented below.

\subsection{Solver}

This optimization model was solved with Gurobi Optimizer version 9.5.1 \cite{gurobi}. The hard-stopping criterion is a mix-integer programming (MIP) optimality gap (that is, the gap between the primal objective bound and the dual objective bound) of 5\%.

\section{Results and Sensitivity Analysis}\label{sec:result}

This section presents the findings of the study. First, we discuss the overall station sizing under different battery capacities from a system design perspective (Section \ref{sec:size1}). Next, we compare various battery sizes in terms of system efficiency (Section \ref{sec:efficiency}). A cost-benefit analysis, integrating transportation efficiency and battery utilization efficiency, is conducted in Section \ref{sec:cost}. Finally, Sections \ref{sec:CO2} and \ref{sec:as} explore \rev{station-level} potential to offer additional services beyond battery swapping, including their carbon reduction capabilities and ability to provide grid services. \rev{We ran extensive analyses on different carbon cost scenarios, also examining multiple days throughout 2023 to capture seasonal variations for a single station.}

\subsection{Station Sizing and Battery Sizing}\label{sec:size1}

We first solve the decision problem described in Section \ref{sec:formulation} to understand the BSS system design in California. 
Each station in California has been sized based on the average daily demand of the NHTS data from the annual truck trips in year 2023. 

\begin{figure}[ht]
\centering
\includegraphics[width=\linewidth,trim={0cm 0cm 0cm 3cm},clip]{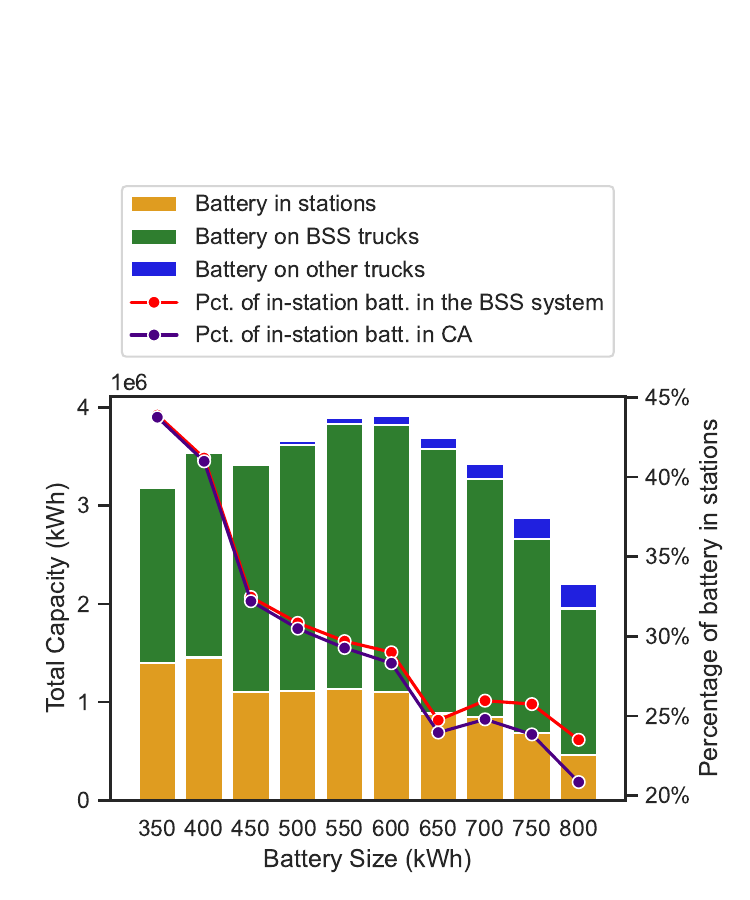}
\caption{Total kWh battery capacity in the battery swapping stations (orange), total kWh battery capacity on trucks using battery swapping systems (green), and total kWh battery capacity for other electrified heavy-duty trucks in California (blue). } \label{fig:num_bat}
\end{figure}

Figure \ref{fig:num_bat} summarizes the total capacity of batteries needed with respect to different battery sizes. Notice that there are three types of batteries, batteries in the battery-swapping stations (orange), batteries on the trucks that utilize BSS services (BSS trucks, green), and batteries in other trucks not using BSS nor needing en-route charging (other trucks, blue). We refer to the trucks that use BSS services and the stations as the BSS system, and the red curve represents the percentage of in-station batteries in this system. The purple curve represents the percentage of batteries in the station compared to all trucks and BSSs in CA.

There are several interesting trends. First, the total capacity of batteries on ``other trucks'' that do not need en-route charging increases, as the proportion of trips that can be completed without recharging increases. In addition, the total capacity of batteries in the swapping stations decreases. This is an intuitive consequence of the fact that larger battery packs for trucks also result in a higher percentage of charging occurring at the depot, with a corresponding reduction in the percentage of charging occurring by swapping. Meanwhile, the total capacity of batteries on BSS trucks initially rises in tandem with the introduction of larger batteries, but subsequently declines in step with the reduction in the number of BSS trucks. Ultimately, the total battery capacity within the BSS system and the overall total first increases and then decreases due to the interplay of these multiple factors.

From the standpoint of the system, including trucks using battery swapping services and stations, 23.5\% to 43.9\% of the battery capacity is located in the stations, while the remaining capacity is distributed among the trucks. From the perspective of the system encompassing all trucks in California and the associated swapping stations, 20.8\% to 43.8\% of the battery capacity is located in the stations.

In such a BSS system, two key considerations are the total capacity of the system and the percentage of battery capacity in stations. It is preferable to have the total capacity as low as possible. With a lower total battery capacity in the system, less batteries are needed, which reduces the investment of such a system. 
For the latter, the energy-carrying capability of trucks is contingent upon battery capacity on the truck, which in turn determines the number of en-route charging events that are required. Therefore, the overall efficiency of the fleet can be enhanced by having more battery capacity on the truck. 
Nevertheless, a greater number of batteries at the station enables the provision of a wider range of services beyond e-mobility. 
Section \ref{sec:efficiency} will provide a more detailed analysis of the trade-off in determining the optimal battery size for the BSS system, with a comparison to the FCS system.



\begin{figure}[htp]
  \centering
  \includegraphics[width=0.8\linewidth]{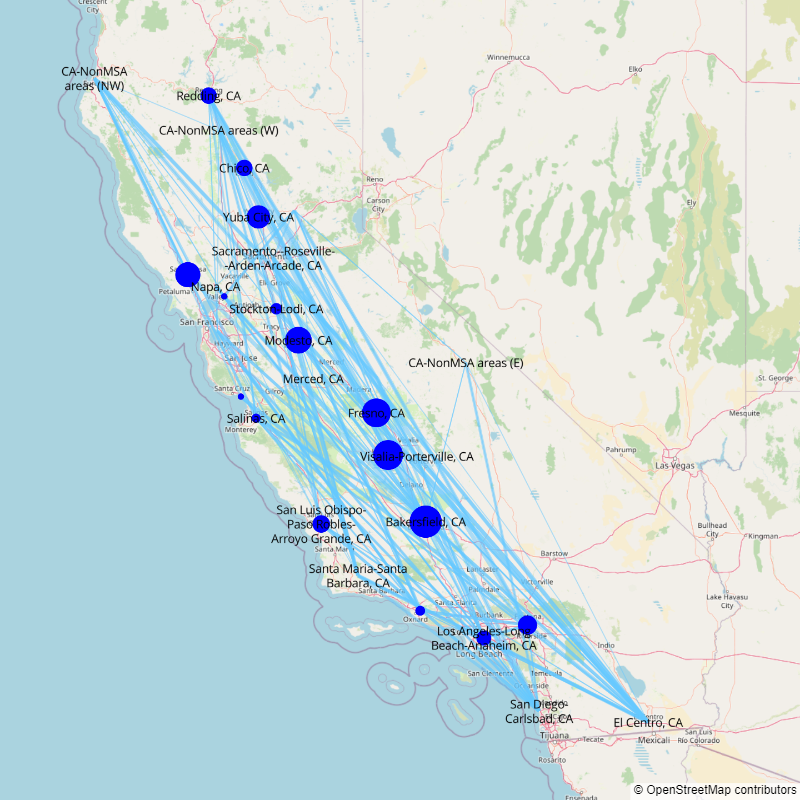}
  \caption{The distribution of batteries in the California intra-trip battery swapping system using 500 kWh batteries. The size of the node represents the log number of batteries present at each station.}\label{fig:full_net}
\end{figure}

\rev{For illustrative purposes, we also present the sizing of the station using a 500 kWh battery size as an example. Figure \ref{fig:full_net} demonstrates the geographical distribution of these batteries. As only intra-zone trips are considered, the largest charging depots are situated in California's Central Valley. }

\subsection{Battery Sizing and System Efficiency}\label{sec:efficiency}

In determining the battery size for the BSS and FCS systems, there are two important factors: time and material. Specifically, we analyze the best battery sizing in terms of transportation efficiency in ton-miles per hour, which examines the potential time savings of battery swapping versus fast charging. We also analyze battery utilization efficiency in ton-mile per kWh, which examines the potential material/cost savings of battery swapping versus fast charging. 

\subsubsection{BSS vs FCS in Transportation Efficiency}

Time is one of the biggest challenges in electrifying supply chains and logistics, especially for long-haul heavy-duty trucks. Liimatainen et. al., for example, have shown that the complexity of routing and scheduling increases with longer refueling times \cite{liimatainen2019}. 

To measure transportation efficiency, we examine the metric of ton-miles per hour. A ton-mile is commonly used in supply chain and logistics for the measurement of the overall level of activity in the economy. Similarly, revenue ton-mile is used in the transportation industry as a determinant of profit. The ton-mile-per-hour metric we defined measures the relative efficiency of supply chain transportation and can be further translated to the labor required per ton-mile of cargo transportation.

\begin{figure}[htbp]
\centering
\includegraphics[width=\linewidth]{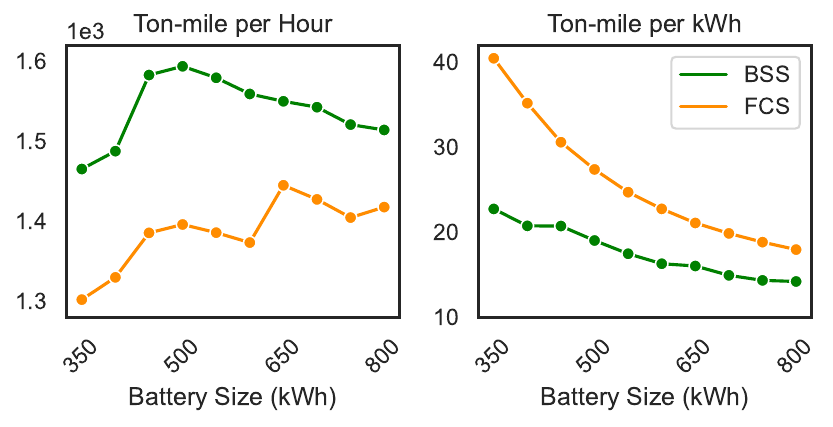}
\caption{Transportation efficiency (left) and battery utilization efficiency (right) analysis comparing different battery sizes and two charging methods.}
\label{fig:effi1}
\end{figure}

In both systems, transportation efficiency depends on the load-carrying capacity and the number and frequency of en-route charging events. When fully loaded, a truck with a 350 kWh battery can carry 10.4\% more cargo mass than a truck with an 800 kWh battery. A charging event occurs when the driving distance exceeds the battery range to enable continued delivery. If the total time exceeds the driver's maximum hours of service, a 10-hour uninterrupted rest period is required. 

As seen in Fig. \ref{fig:effi1} left subfigure, the BSS system exhibits greater transportation efficiency relative to FCS due to the faster service time in recharging. For FCS, the charging and resting events have a significant impact on efficiency. As the battery size increases to 650 kWh, there is a noticeable increase in transport efficiency and less resting time is required for some trips. In the context of BSS, the optimal battery pack size for transport efficiency is 450 to 500 kWh. This size does not require the battery to be recharged frequently, yet also does not occupy excessive weight or cargo space. The optimal battery size for a given vehicle, however, is a function of trade-offs between various key performance indicators, such as transport efficiency and battery utilization efficiency.

In summary, trucks operating on FCS require larger batteries to increase transportation efficiency. Larger batteries reduce charging frequency, yet also maintain operational tasks within the hours-of-service limits. BSS systems are better suited for trucks with medium-sized batteries, allowing for increased cargo capacity.

\subsubsection{BSS vs FCS in Battery Utilization Efficiency}

Battery material requirements are a crucial issue, especially considering the complex economic and political supply chain issues. We examine the number of excess batteries required for the swapping service and those needed for a more conventional FCS setup. For BSS, we naturally expect that more batteries are needed to electrify the system. The metric, ``ton-mile per total kWh capacity'', is analyzed to measure battery utilization efficiency\footnote{In this analysis, we consider ton-miles in the context of daily operational usage, which is reflective of the battery utilization profile throughout the product's life cycle.}. 

In the case of FCS, 100\% of the batteries are located on the trucks. In Fig. \ref{fig:effi1} right subfigure, for a battery of the same size, the battery utilization efficiency is consistently higher for FCS than for BSS. Nevertheless, it can be observed that a smaller battery always has a higher battery utilization efficiency. Suppose we examine the battery sizes that optimize transportation efficiency for BSS and FCS (450-500 kWh and 650 kWh, respectively). BSS systems with battery capacities of 450-500 kWh have a battery utilization efficiency that is only 2.9\%, and 10.8\% lower than that observed in FCS systems with a battery capacity of 650 kWh. 

A comparable analysis is presented in the conference paper cited as \cite{rtw_ACC}, which employs a rather optimistic setting based on a single-station analysis. This paper reaches the same general conclusion with a more realistic demand data set, a detailed network configuration, and a more comprehensive analysis of system sizing. It should also be noted that idling batteries in a station may not necessarily be a waste of resources. These batteries can be used to provide multiple grid services, similar to battery storage systems \cite{richard2018a}. In section \ref{sec:as}, we examine BSS's ability to provide grid services. 

In conclusion, the potential to utilize smaller battery sizes increases battery utilization efficiency for BSS, thereby negating the necessity to purchase and install additional batteries.

\subsection{Cost-Benefit Analysis}\label{sec:cost}

\rev{Transportation efficiency and battery utilization efficiency offer two distinct aspects for evaluating the performance of BSS from a systems perspective. These findings are fundamental, without assuming a specific business model or competition between different stakeholders. 
However, while both time and battery material are crucial factors, we must select a trade-off for the optimal battery size. In this section, we present an approach to assess this trade-off from an economic perspective, translating time into labor hours and battery capacity into upfront investment costs. }

\rev{Based on a market survey, the average price per lithium-ion battery pack was \$151/kWh in 2022 \cite{2022bloomberg}. The average truck driver salary in California is \$28.02 per hour in May 2023 \cite{driver_wage}. We additionally assume these batteries have a lifetime of eight years}\footnote{\rev{Based on the studied daily operational usage profile, the equivalent discharge cycles are calculated from daily energy consumption relative to the total battery capacity in the system. BSS averages 1.04 cycles per day and FCS 1.52 cycles per day. The actual battery lifetime depends on system operational settings. However, assuming 260 workdays per year and a total battery lifespan of 2000 full cycles \cite{cunanan2021, moultak2017}, the estimated battery lifespan is 7.41 years for BSS and 5.07 years for FCS.}}. 

\begin{figure}[htpb]
\centering
\includegraphics[width=\linewidth]{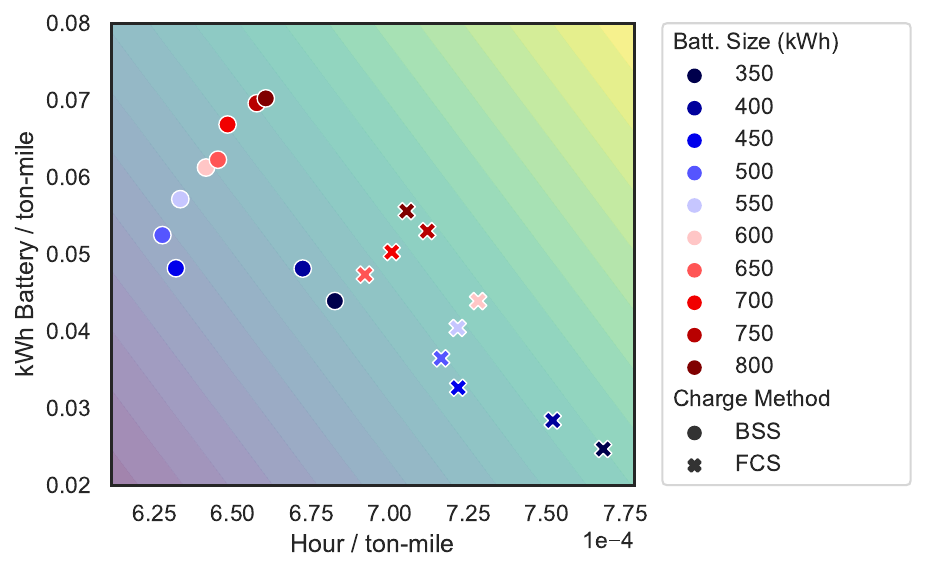}
\caption{Determining optimal battery sizes through cost-benefit analysis on transportation efficiency and battery utilization efficiency. The background isolines illustrate the optimal balance where the marginal cost of additional battery capacity intersects with the increased labor costs.}
\label{fig:compare}
\end{figure}

\rev{In order to ascertain the optimal allocation of resources, we conducted a cost-benefit analysis, comparing the marginal economic cost of purchasing an additional kWh of battery capacity against the increased labor cost resulting from longer transportation times. The background isolines of Fig. \ref{fig:compare} illustrate the optimal balance, where the marginal cost of additional battery capacity intersects with the increased labor costs. The lower left direction represents the most efficient scenario. The results in Fig. \ref{fig:compare} demonstrate that BSS systems exhibited significantly greater efficiency than FCS systems with respect to transportation efficiency and battery utilization efficiency\footnote{It is important to note that the initial investment required for the implementation of a swapping system is not taken into account in this analysis. }. }

\rev{The optimal battery sizes for both systems are invariably sensitive to the cost of batteries, driver costs, and the assumed battery life-span. With our assumptions in Fig. \ref{fig:compare}, the most efficient battery size for a BSS is 450 kWh, followed by 500 kWh. For FCS, it is 650 kWh, followed by 450 kWh.}

\rev{Considering the uncertainty in battery acquisition costs and labor costs, we assume these factors to be variable. Given the ongoing trend of decreasing battery costs and increasing battery lifespan, coupled with the concomitant increase in driver wages and benefits, the background isolines are expected to be steeper. In this scenario, a per-unit increase in travel time becomes more costly, while a per-unit increase in battery assets becomes less costly. In the extreme case where the background isolines are vertical, indicating indifference to battery cost, BSS systems would still prefer sizes of 500 kWh and 450 kWh, while FCS would favor larger sizes such as 650 kWh and 700 kWh.}

\rev{Conversely, the background isolines may flatten with the gradual introduction of autonomous driving into logistics systems, leading to a significant reduction in the cost per hour of ``labor''. Meanwhile, ever-changing government policies related to battery supply chains and the cost of critical battery materials could drive battery prices upward. If the background isolines are horizontal--meaning vehicle operation time is no longer a concern--the cost of both systems would be insensitive to the increase in travel time, but very sensitive to the total battery capacity in the system. A smaller battery size of 350 kWh will become more favorable for both BSS and FCS systems.}

\subsection{Carbon Aware Operation}\label{sec:CO2}


\begin{figure*}[hbt]
\centering
\includegraphics[width=0.8\linewidth]{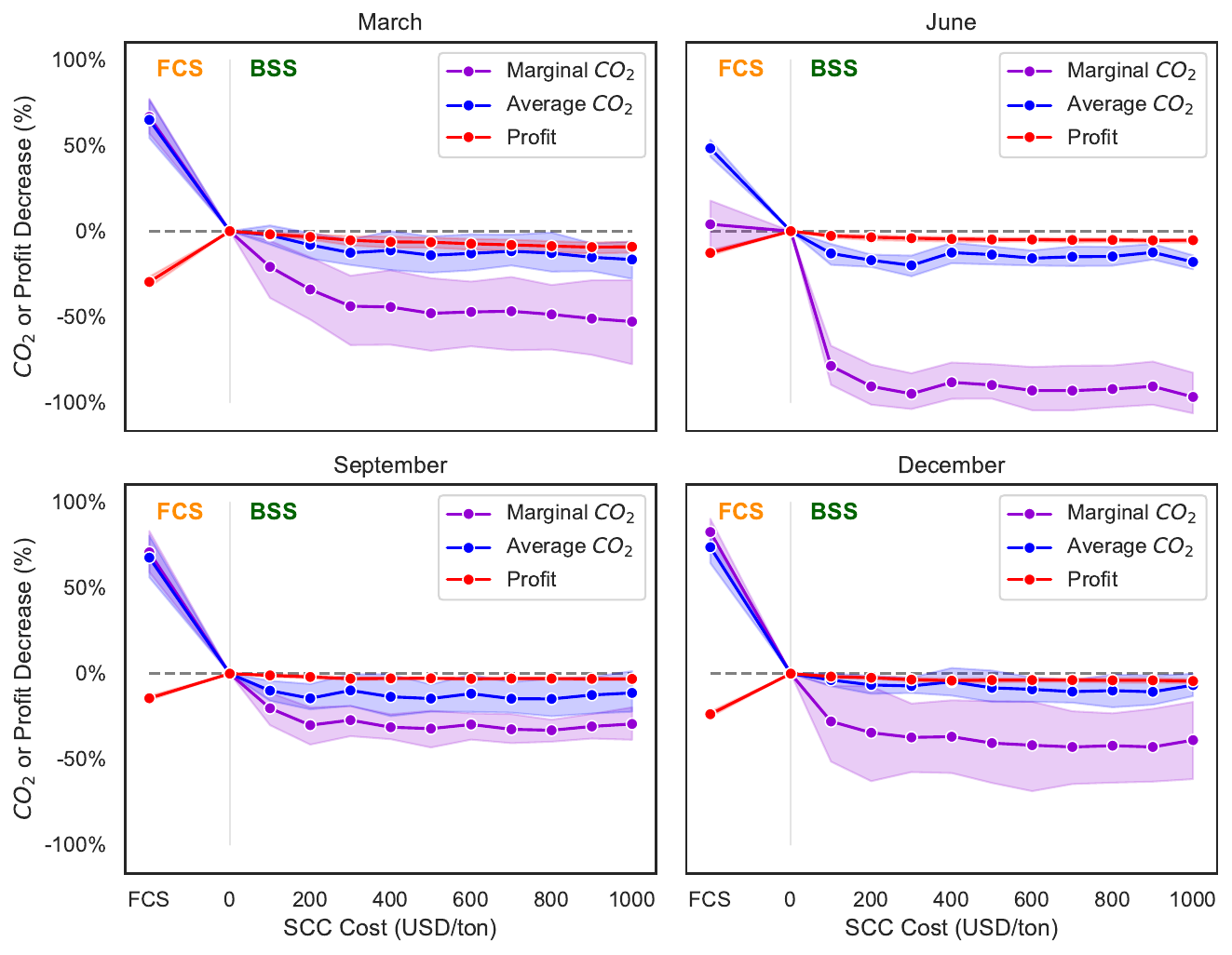}
\caption{$\text{CO}_2$ emissions and profits change in percentage comparing FCS and BSS with different social carbon costs, under a strategy that minimizes marginal $\text{CO}_2$ emissions. The reference point for the percentage calculation is BSS with zero carbon costs. }
\label{fig:co2_4mo}
\end{figure*}

We further examine the potential for BSS to reduce carbon emissions by shifting charging loads over time. BSS can provide arbitrage opportunities in energy markets, which could be a significant source of profit. Considering a potential carbon cost would help to regulate the charging and discharging behavior in a carbon-aware manner. 
\rev{To incorporate this aspect, the objective function has the addition term from \eqref{obj_operation} to account for SCC, where $\xi_t^{\text{CO}_2}$ is the emission factor -- in particular, we tested the average emission factor $\xi_t^{\text{Avg}}$ and marginal emission factor $\xi_t^{\text{Mar}}$. }

\rev{\begin{align}\label{scc}
\begin{split}
\min \quad &  J_\text{op} + \sum_{t\in \mathcal{T}} 
SCC \xi_t^{\text{CO}_2}  e^{\text{G}}_t 
\end{split}
\end{align}}

This analysis is based on station-level operations. For this analysis, we extracted characteristic customer demand profiles for each station. Figure \ref{fig:co2_4mo} illustrates the change of $\text{CO}_2$ emissions and daily operation profits of the studied station \rev{(single station)} in different scenarios. 

The baseline scenario uses battery swapping as the charging mechanism and assumes no carbon costs (SCC = 0 USD/ton). We compared it to fast charging, which has no load-shifting capability, and battery swapping with different levels of SCC. The profits indicated in the results solely encompass operational costs and revenues, without the costs associated with carbon. 
For FCS, customers are always charged immediately upon arrival. The energy demand transfer to the grid and total emissions are calculated using the $\text{CO}_2$ emission factors at the moment. Meanwhile, BSS is capable of providing energy arbitrage through shifting load profiles, and decouple the timing of customer demand for charging and the demand for energy from the grid.
The mean marginal $\text{CO}_2$ emissions (purple curve) of FCS range from 4\% to 82\% higher than those of BSS with zero SCC. This is attributable to the inherently volatile nature of marginal emission factors. The higher emissions from FCS are more obvious using average emission signals. FCS has 39\% to 92\% higher emissions compared to BSS. Additionally, the daily operational profit of FCS is observed to be 9.9\% to 34\% lower than that of BSS. The increase in profits for the battery swapping system is because BSS is capable of providing energy arbitrage. The change in $\text{CO}_2$ emissions depends on the relationship between carbon emission signals and price signals from the grid. It is important to acknowledge the unpredictability of the relationship between electricity prices and carbon emission factors. 
However, in this case, the provision of energy arbitrage by BSS has the additional benefit of reducing $\text{CO}_2$ emissions in comparison to FCS. 

With the carbon-aware feature in BSS operation, $\text{CO}_2$ emissions and net profits under different SCCs are compared. Minimizing marginal emissions was selected as the objective because it usually effectively reduces average emissions as well. The results in Fig. \ref{fig:co2_4mo} show that it is possible to reduce marginal $\text{CO}_2$ emissions by 20-96\%  (51\% on average), average $\text{CO}_2$ emissions by 2.4-20\%  (12\% on average), with only a 1.1-9.3\% reduction in profits (4.3\% on average). With the given pricing mechanism, the system never violates the customer's requirements, but it does change the in-station battery charge and discharge profiles to include SCC. 

The potential for $\text{CO}_2$ reduction varies across different days and different months of the year. However, in general, a significant reduction in $\text{CO}_2$ emissions can be achieved with only a minor loss in profit. Such a benefit exists when the SCC is a positive value in the context of carbon-aware optimization. This feature serves to underscore the potential for carbon reduction associated with BSS in comparison to FCS. It also highlights the significance of incorporating carbon-aware management into the operational framework of BSS systems.







\subsection{Grid Ancillary Services}\label{sec:as}

\rev{In California, due to the high penetration of renewable energy, the grid experiences significant fluctuations in supply and demand. As a result, there is a continuous need for ancillary services such as frequency regulation and demand response to maintain grid stability. BSS and electrified truck fleets, with their large battery capacities and flexible charging schedules, present an opportunity to provide these services, offering additional revenue streams while enhancing grid reliability.}

Existing literature has already shown that stationary batteries in fast charging stations can be used to provide grid services \cite{richard2018a}. The in-station battery capacities shown in Fig. \ref{fig:num_bat} can be used to provide energy arbitrage and ancillary services similarly. 
In the context of BSS, this approach enables the battery to be treated as an asset, similar to stationary battery storage. Additionally, it facilitates the provision of swapping as a service, along with other services that battery storage systems are capable of offering. 

We consider a BSS that provides frequency regulation service to the grid by participating in the day ahead market to provide regulation up (RU) and regulation down (RD). \rev{While relying solely on day-ahead prices may lead to inefficiencies due to real-time price deviations, we adopt this approach to illustrate the operational and economic implications of BSS participation in structured electricity markets.} We only account for the capacity payment portion of the compensation, which is based on the actual service provided to CAISO. 

\rev{The objective function has the addition terms to account for AS, where $RU^{\text{G}}_t$ and $RD^{\text{G}}_t$ are historical clearing prices of the ancillary service DAM market in California. }

\rev{\begin{align}\label{as}
\begin{split}
\min \quad &  J_\text{op} + \sum_{t\in \mathcal{T}} 
(RU^{\text{G}}_t + RD^{\text{G}}_t)  e^{\text{G}}_t 
\end{split}
\end{align}}

\begin{figure}[htb]
\centering
\includegraphics[width=\linewidth]{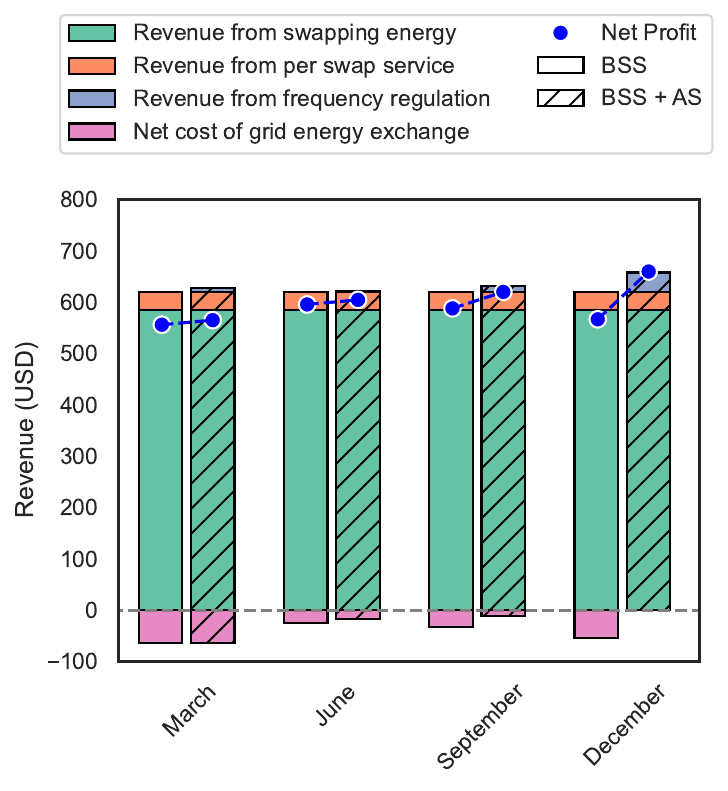}
\caption{Changes in profitability between cases without \rev{(left) and with (right)} the provision of ancillary services. \rev{The result is based on a single station and shows the average of total net profit over seven days in the first week, over the variation of electricity prices and ancillary service prices for each month studied.}}

\label{fig:as}
\end{figure}

\rev{We studied the first week of March, June, September, and December, and Fig. \ref{fig:as} shows the average values of each month.} 

\rev{Based on the studied station,} we compared the breakdown of daily operation profit with (dark shade) and without (light shade) providing grid frequency regulation across four months in Fig. \ref{fig:as}. The net profit increases from 1.3\% in June to 16.9\% in December. 
While there are day-to-day variations, the general need for regulation-down is higher in California during the winter. One reason for this is that, during the winter months, electricity demand is typically lower than that during the summer due to the reduced need for air conditioning. This reduction in demand can result in an oversupply of electricity, increasing the need of RD. Even a naive bidding strategy to provide regulation down would benefit the system.

In conclusion, the provision of ancillary services varies greatly depending on location, signal, and market clearing prices. We find that providing RD is generally advantageous, and other ancillary services may not be economically significant. For instance, during the studied period in 2023, only 0.9\% of the observed hours showed RU prices surpassing those of the wholesale market. Consequently, the system never provides regulation up, and the economic benefits come solely from regulation down. Thus, while BSS can serve as an ancillary service provider, a complex bidding strategy is unnecessary. The system can benefit from a simple approach that provides regulation down at maximum capacity. 

\section{Limitations}

\rev{Our comparative analysis of BSS versus FCS represents an initial step toward understanding the trade-offs between various aspects of system efficiency, though we acknowledge the limitations of our deterministic approach. The current model does not account for uncertainties in truck arrival patterns or operational contingencies that would be present in real-world freight networks. }

\rev{In addition, our model employs a simplified battery model that does not fully capture the complexities of battery behavior, including non-linear charging efficiency curves and degradation mechanisms. A more sophisticated battery model would provide insights into how the different operational patterns of BSS (with more frequent but shallower cycles) versus FCS (with deeper but less frequent cycles) affect long-term battery health and replacement schedules. This is particularly a promising direction for heavy-duty applications where the high energy throughput and operational demands may accelerate degradation mechanisms beyond what is typically seen in light-duty vehicles.}

\rev{Another limitation is that our study focuses on operational aspects, and we did not perform a comprehensive CAPEX analysis due to the significant uncertainties and variations in infrastructure costs. The charger cost is highly dependent on three key factors: charging capacity, location characteristics, and grid connection requirements. Recent literature provides varying cost estimates for charging infrastructure. For reference of the readers, the CAPEX of charging stations is studied in \cite{zhu2023, michaelnicholas2019, gamage2023}. These three papers’ estimations of charger cost differ by up to a factor of 20. Table \ref{tab:char_cost} summarizes our compiled estimates for 150 kW and 450 kW chargers based on these sources, providing a range that reflects both current market conditions and regional variations.
While there is insufficient market data for a fair estimation of automated swapping equipment costs, as this technology is still emerging in the heavy-duty vehicle sector, a comprehensive sensitivity analysis to account for this uncertainty would be very valuable to the industry. 
}

\begin{table*}[]
    \centering
    \rev{\caption{Comparison of Charger Cost Estimates from Literature}\label{tab:char_cost}
    \begin{tabular}{cccc}
    \toprule
    Literature & 450 kW Charger & 150 kW Charger & Notes \\
    \hline
    Zhu et al., 2024 \cite{zhu2023} & \$91,799 & \$31,481 & Price in RMB (based on China) \\
    Gamage et al., 2023 \cite{gamage2023} & \$1,875,195 & \$625,065 & Price based on CA highway corridor \\
    Nicholas, 2019 \cite{michaelnicholas2019} & \$225,000 & \$75,000 & International Council on Clean Transportation \\
    \bottomrule
    \end{tabular}}
\end{table*}

\section{Conclusion}\label{sec:conclu}


\rev{This paper presents a comprehensive evaluation of charging infrastructure strategies for electric heavy-duty trucks, developing an integrated optimization framework for BSS planning and operations. Using California’s freight network as a case study, we systematically compared BSS and FCS across multiple performance dimensions while exploring the dual-use potential of stationary batteries for grid services.}

Our analysis revealed that BSS with medium-sized batteries \rev{offers} optimal overall efficiency in terms of transportation and battery utilization. In contrast, trucks that rely on FCS require larger batteries to offset the impact of extended charging times.
\rev{We also investigated how optimal battery sizing responds to evolving market conditions. Our sensitivity analysis shows that declining battery costs coupled with rising labor expenses favor smaller on-board batteries with more frequent swapping. Conversely, scenarios with increasing battery costs (due to raw material constraints) and decreasing labor costs (from autonomous driving adoption) shift the optimum toward larger batteries with fewer swaps.}
\rev{Beyond operational efficiency,} our highlighted BSS's \rev{substantial} potential for \rev{reducing} $\text{CO}_2$ emission and \rev{enhanced} profitability through energy arbitrage and grid ancillary services. 

These findings underscore the importance of integrating BSS into future electric truck charging networks and adopting carbon-aware operational frameworks. \rev{As electrification of heavy-duty transportation accelerates, the intelligent deployment and operation of charging infrastructure will be critical to achieving both economic viability and environmental sustainability goals.}

\section{Acknowledgement}
This work is supported, in part, by TotalEnergies OneTech under award No. 048589-001 and the National Science Foundation AI Institute for Optimization (AI4Opt) under
Award No. 2112533 at the University of California, Berkeley, USA.

\appendices

\section{Energy Estimation}\label{energy_model}

\rev{
We use the following longitudinal vehicle dynamics model from literature \cite{basso2019}. For simplicity, we assume that the speed profile of trucks is constant acceleration up to maximum speed, maintaining the maximum speed, and constant deceleration until the vehicle stops. In addition, the net elevation change of the route is used to calculate the average road grade. }

\begin{equation}\label{energy1}
    ma(t) = F_t (t) - (F_g(t) + F_r(t) + F_a(t))
\end{equation}
\begin{align}
    \begin{split}
        F_g(t) = mg\sin\theta(t), \\
        F_r(t) = mgC_r\cos\theta(t), \\
        F_a(t) = 0.5C_dA\rho v(t)^2
    \end{split}
\end{align}

\begin{equation}
    E = E_a + E_c + E_d
\end{equation}

\begin{align}
    \begin{split}
        E_a = \frac{d_a(m\bar a_a + mg\sin\bar\theta + mgC_r\cos\bar\theta+ 0.5C_dA\rho\bar v^2/2}{3600\eta_a} \\
        E_c = \frac{(d - d_a - d_d)(mg\sin\bar\theta + mgC_r\cos\bar\theta+ 0.5C_dA\rho\bar v^2}{3600\eta_c} \\
        E_d = \frac{d_d(m\bar a_d + mg\sin\bar\theta + mgC_r\cos\bar\theta+ 0.5C_dA\rho\bar v^2/2}{3600\eta_d} 
    \end{split}
\end{align}

\rev{Table \ref{tab:para_energy} shows the selection of parameters and their meaning. }




\begin{table}[h]
\centering
\rev{
    \caption{Parameter values for energy estimations}\label{tab:para_energy}
    \begin{tabular}{ccc}
    \toprule
     Notion    & Meaning &  Value\\
     \hline
       $m$  &  Vehicle mass (kg) &  37194.6 \\
       $F_g$  &  Gravity force (N) &  9.81\\
       $a$  & Instantaneous acceleration (m/s$^2$) & 1 \\
       $V_{max}$  & Maximum speed (km/h) & 80 \\
       $C_r$ & Rolling resistance coefficient &  0.0061 \\
       $C_d$ & Drag coefficient & 0.581 \\
       $A$ & Frontal surface area of the vehicle (m$^2$) & 10.0684 \\
       $\rho$ & Air density (kg/m$^3$) & 1.225 \\
       $\eta_a$ & Powertrain efficiency when accelerating & 0.9 \\
       $\eta_c$ & Powertrain efficiency during constant speed & 0.9 \\
       $\eta_d$ & Powertrain efficiency when decelerating & 0.9 \\
       $F_t$  & Force generated by the powertrain (N) &  /\\
       $F_r$  & Rolling friction force (N) &  /\\
       $F_a$  & Aerodynamic friction force (N) & / \\
       $\theta$ & Instantaneous road inclination angle &  /\\
    \bottomrule
    \end{tabular}
    \label{tab:my_label}
    }
\end{table}

\section{Network Decomposition}\label{network_decom}
\begin{figure*}[htb]
    \centering
    \subfloat[Original distribution of the edge visit frequency in the network.]{\includegraphics[height=0.2\linewidth]{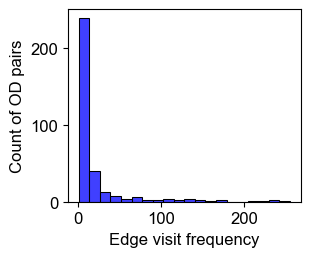}} \quad \quad 
    \subfloat[Distributions of the edge visit frequencies in the decomposed networks after scaling with $\beta_1 = 30$ and $\beta_2 = 5$.]{\includegraphics[height=0.26\linewidth]{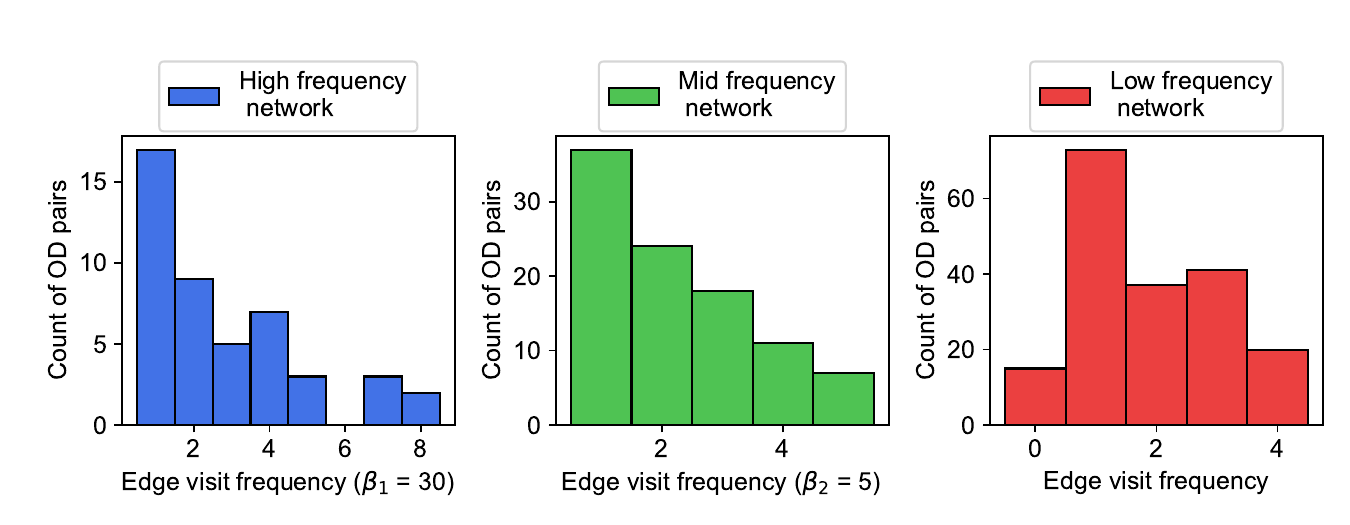}}
    
    \caption{Demand network decomposition with $\beta_1 = 30$, $\beta_2 = 5$. We show the original distribution of the edge visit frequency and updated edge visit frequencies in high-, medium-, and low-frequency networks. }
    \label{fig:dist}
\end{figure*}

\rev{
There are an average of 6236 long-haul truck trips per day across the network based on the NHTS datasets. We decompose the network into high-, medium-, and low-frequency networks to solve the problem efficiently. Two coefficients $\beta_1$ and $\beta_2$ are selected to decompose the network into: 1) a high-frequency network, consisting of all the edges with a daily trip frequency greater than $\beta_1$; 2) a medium-frequency network, with a daily trip frequency between $\beta_1$ and $\beta_2$; and 3) a low-frequency network, with a daily trip frequency less than $\beta_2$. We scale the edge visit frequency in the high-frequency network by $\beta_1$ and medium-frequency network by $\beta_2$, after which the edge visit frequencies are rounded to integers. The two coefficients $\beta_1$ and $\beta_2$ are chosen so that the three networks have relatively the same size after scaling. Figure \ref{fig:dist} shows the set of coefficients used in the paper. 
}

\section{\rev{Sensitive analysis on Different Charging Rates}}\label{char_speed}

\rev{Section \ref{sec:efficiency} and \ref{sec:cost} are based on the assumption that FCS requires one hour for recharge (roughly 1C, accounting for partial charging cycles and operational overhead) and BSS takes 30 minutes to complete a swap. To address concerns about these potentially conservative estimates, we conducted sensitivity analyses with progressively faster-charging scenarios.}

\rev{We tested three alternative charging profiles: (1) 30-minute stay time at FCS (roughly 2C) and 15-minute stay time at BSS; (2) 20-minute stay time at FCS (roughly 3C) and 10-minute stay time at BSS; and (3) 15-minute stay time at FCS (roughly 5C) and 15-minute stay time at BSS. These scenarios align with the industry trajectory toward Megawatt Charging Systems for heavy-duty applications and advanced battery-swapping technologies while maintaining the relative time advantage of BSS over FCS observed in current implementations.}

\begin{figure*}[htbp]
\centering
\subfloat[Profile 1: 30-min FCS and 15-min BSS]
{\includegraphics[height=0.28\linewidth, clip,trim=0 0 50mm 0]{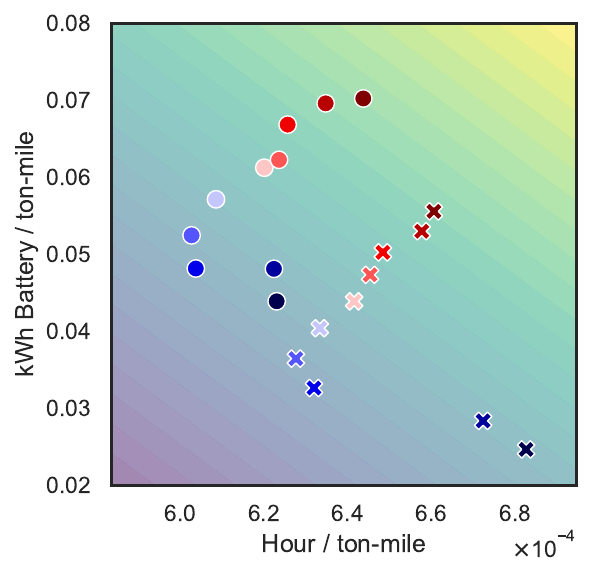}}
\subfloat[Profile 2: 20-min FCS and 10-min BSS]
{\includegraphics[height=0.28\linewidth, clip,trim=0 0 50mm 0]{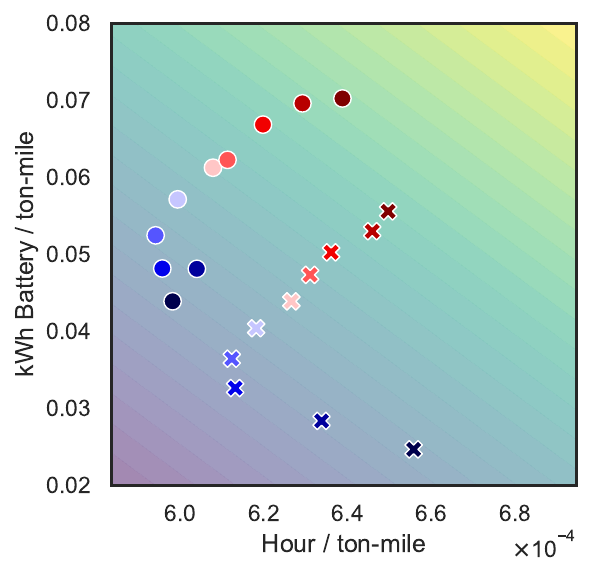}}
\subfloat[Profile 3: 15-min FCS and 15-min BSS]
{\includegraphics[height=0.28\linewidth, clip,trim=0 0 50mm 0]{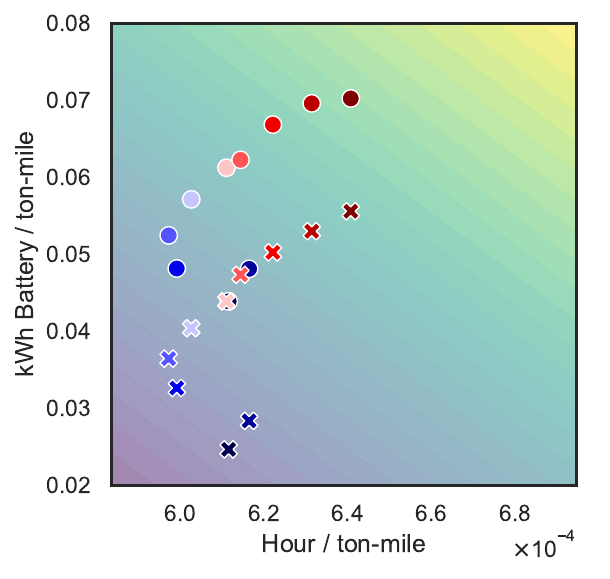}}
{\includegraphics[height=0.28\linewidth, clip,trim=100mm 0 0mm 0]{fig/appdC/color_map_0.2_0.2.pdf}}
\caption{Cost-benefit analysis of optimal battery sizes in different alternative charging profiles. Partial charge cycles and operational overhead are accounted for in the charge time. }
\label{fig:appC}
\end{figure*}

\rev{Figure \ref{fig:appC} shows the results of these sensitivity analyses. While the analytical methods presented in Sections \ref{sec:efficiency} and \ref{sec:cost} remain unchanged, the outcomes vary in different scenarios. }

\rev{A consistent pattern emerges: regardless of charging speed assumptions, BSS continues to favor smaller battery sizes compared to FCS when the swapping process remains faster than FCS recharging. This relationship holds across all tested scenarios, reinforcing the fundamental trade-off between battery capacity and charging infrastructure.}

\rev{Interestingly, when both systems require equal time for an en-route charging stop and we consider only battery acquisition costs while disregarding infrastructure investments, FCS consistently demonstrates cost advantages over BSS due to its lower total battery requirements.}

\rev{However, higher charging rates in FCS introduce additional considerations not captured in our model, including potential battery degradation effects, thermal management challenges, and significantly higher peak power demands on the electrical grid. These factors may limit the practical implementation of very high charging rates in real-world operations, particularly for heavy-duty applications.}

\bibliographystyle{IEEEtran}  %
\bibliography{refs}

\providecommand{\noopsort}[1]{}
\begin{thebibliography}{10}
\providecommand{\url}[1]{#1}
\csname url@samestyle\endcsname
\providecommand{\newblock}{\relax}
\providecommand{\bibinfo}[2]{#2}
\providecommand{\BIBentrySTDinterwordspacing}{\spaceskip=0pt\relax}
\providecommand{\BIBentryALTinterwordstretchfactor}{4}
\providecommand{\BIBentryALTinterwordspacing}{\spaceskip=\fontdimen2\font plus
\BIBentryALTinterwordstretchfactor\fontdimen3\font minus \fontdimen4\font\relax}
\providecommand{\BIBforeignlanguage}[2]{{%
\expandafter\ifx\csname l@#1\endcsname\relax
\typeout{** WARNING: IEEEtran.bst: No hyphenation pattern has been}%
\typeout{** loaded for the language `#1'. Using the pattern for}%
\typeout{** the default language instead.}%
\else
\language=\csname l@#1\endcsname
\fi
#2}}
\providecommand{\BIBdecl}{\relax}
\BIBdecl

\bibitem{EPA2023}
{U.S. Environmental Protection Agency}, ``Greenhouse {{Gas Emissions Standards}} for {{Heavy-Duty Vehicles-Phase}} 3,'' Apr. 2023.

\bibitem{EPA2024}
------, ``Final {{Rule}}: {{Greenhouse Gas Emissions Standards}} for {{Heavy-Duty Vehicles}} -- {{Phase}} 3,'' https://www.epa.gov/regulations-emissions-vehicles-and-engines/final-rule-greenhouse-gas-emissions-standards-heavy-duty, Feb. 2024.

\bibitem{liimatainen2019}
H.~Liimatainen, O.~{\noopsort{vliet}}{van Vliet}, and D.~Aplyn, ``The potential of electric trucks -- {{An}} international commodity-level analysis,'' \emph{Applied Energy}, vol. 236, pp. 804--814, Feb. 2019.

\bibitem{wang2024b}
R.~Wang, P.~Keyantuo, T.~Zeng, J.~Sandoval, A.~Vishwanath, H.~Borhan, and S.~Moura, ``Robust routing for a mixed fleet of heavy-duty trucks with pickup and delivery under energy consumption uncertainty,'' \emph{Applied Energy}, vol. 368, p. 123407, Aug. 2024.

\bibitem{cabukoglu2018}
E.~{\c C}abukoglu, G.~Georges, L.~K{\"u}ng, G.~Pareschi, and K.~Boulouchos, ``Battery electric propulsion: {{An}} option for heavy-duty vehicles? {{Results}} from a {{Swiss}} case-study,'' \emph{Transportation Research Part C: Emerging Technologies}, vol.~88, pp. 107--123, Mar. 2018.

\bibitem{pelletier2017}
\BIBentryALTinterwordspacing
S.~Pelletier, O.~Jabali, G.~Laporte, and M.~Veneroni, ``\BIBforeignlanguage{en}{Battery degradation and behaviour for electric vehicles: {Review} and numerical analyses of several models},'' \emph{\BIBforeignlanguage{en}{Transportation Research Part B: Methodological}}, vol. 103, pp. 158--187, Sep. 2017. [Online]. Available: \url{https://www.sciencedirect.com/science/article/pii/S0191261516303794}
\BIBentrySTDinterwordspacing

\bibitem{tomaszewska2019}
\BIBentryALTinterwordspacing
A.~Tomaszewska, Z.~Chu, X.~Feng, S.~O'Kane, X.~Liu, J.~Chen, C.~Ji, E.~Endler, R.~Li, L.~Liu, Y.~Li, S.~Zheng, S.~Vetterlein, M.~Gao, J.~Du, M.~Parkes, M.~Ouyang, M.~Marinescu, G.~Offer, and B.~Wu, ``\BIBforeignlanguage{en}{Lithium-ion battery fast charging: {A} review},'' \emph{\BIBforeignlanguage{en}{eTransportation}}, vol.~1, p. 100011, Aug. 2019. [Online]. Available: \url{https://www.sciencedirect.com/science/article/pii/S2590116819300116}
\BIBentrySTDinterwordspacing

\bibitem{heliox}
{Heliox}, ``Heliox: {{Scalable EV}} charging infrastructure for your e-{{Truck}} fleet,'' https://www.heliox-energy.com/us-solutions/e-truck.

\bibitem{Kempower}
``Power {{Your Electric Fleet}} with {{Kempower Solutions}},'' https://kempower.com/electric-fleets/.

\bibitem{volvo}
``Breakthrough for fast charging of electric trucks -- {{Volvo Trucks}} launches new service,'' https://www.volvotrucks.com/en-en/news-stories/press-releases/2023/oct/breakthrough-for-fast-charging-of-electric-trucks-volvo-trucks-l.html.

\bibitem{biden}
{The White House}, ``{{Biden-Harris Administration Announces New Standards}} and {{Major Progress}} for a {{Made-in-America National Network}} of {{Electric Vehicle Chargers}},'' https://www.whitehouse.gov/briefing-room/statements-releases/2023/02/15/fact-sheet-biden-harris-administration-announces-new-standards-and-major-progress-for-a-made-in-america-national-network-of-electric-vehicle-chargers/, 2023.

\bibitem{li2024}
Y.~Li, F.~Zhu, L.~Li, and M.~Ouyang, ``Electrifying heavy-duty truck through battery swapping,'' \emph{Joule}, May 2024.

\bibitem{chen2022}
X.~Chen, K.~Xing, F.~Ni, Y.~Wu, and Y.~Xia, ``An {{Electric Vehicle Battery-Swapping System}}: {{Concept}}, {{Architectures}}, and {{Implementations}},'' \emph{IEEE Intelligent Transportation Systems Magazine}, vol.~14, no.~5, pp. 175--194, Sep. 2022.

\bibitem{arora2023}
\BIBentryALTinterwordspacing
A.~Arora, M.~Murarka, D.~Rakshit, and S.~Mishra, ``\BIBforeignlanguage{en}{Multiobjective optimal operation strategy for electric vehicle battery swapping station considering battery degradation},'' \emph{\BIBforeignlanguage{en}{Cleaner Energy Systems}}, vol.~4, p. 100048, Apr. 2023. [Online]. Available: \url{https://www.sciencedirect.com/science/article/pii/S2772783122000462}
\BIBentrySTDinterwordspacing

\bibitem{china2023}
CAERI, ``Annual {{Report}} on the {{Development}} of {{China}}'s {{Commercial Vehicle Industry}},'' {China Automotive Engineering Research Institute Co., Ltd.} and {China Association of Automobile Manufacturers FAW Jiefang Group Co., Ltd.}, Tech. Rep., 2023.

\bibitem{zhang2024a}
Y.~Zhang, X.~Wang, and B.~Zhi, ``Strategic investment in electric vehicle charging service: {{Fast}} charging or battery swapping,'' \emph{International Journal of Production Economics}, vol. 268, p. 109136, Feb. 2024.

\bibitem{zhu2023}
F.~Zhu, L.~Li, Y.~Li, K.~Li, L.~Lu, X.~Han, J.~Du, and M.~Ouyang, ``Does the battery swapping energy supply mode have better economic potential for electric heavy-duty trucks?'' \emph{eTransportation}, vol.~15, p. 100215, Jan. 2023.

\bibitem{wu2021a}
X.~Wu, P.~Liu, and X.~Lu, ``Study on {{Operating Cost Economy}} of {{Battery-Swapping Heavy-Duty Truck}} in {{China}},'' \emph{World Electric Vehicle Journal}, vol.~12, no.~3, p. 144, Sep. 2021.

\bibitem{yang2018a}
L.~Yang, C.~Hao, and Y.~Chai, ``Life {{Cycle Assessment}} of {{Commercial Delivery Trucks}}: {{Diesel}}, {{Plug-In Electric}}, and {{Battery-Swap Electric}},'' \emph{Sustainability}, vol.~10, no.~12, p. 4547, Dec. 2018.

\bibitem{liu2019}
X.~Liu, T.~Zhao, S.~Yao, C.~B. Soh, and P.~Wang, ``Distributed {{Operation Management}} of {{Battery Swapping-Charging Systems}},'' \emph{IEEE Transactions on Smart Grid}, vol.~10, no.~5, pp. 5320--5333, Sep. 2019.

\bibitem{ban2019a}
M.~Ban, J.~Yu, M.~Shahidehpour, D.~Guo, and Y.~Yao, ``Electric {{Vehicle Battery Swapping-Charging System}} in {{Power Generation Scheduling}} for {{Managing Ambient Air Quality}} and {{Human Health Conditions}},'' \emph{IEEE Transactions on Smart Grid}, vol.~10, no.~6, pp. 6812--6825, Nov. 2019.

\bibitem{tan2019}
X.~Tan, G.~Qu, B.~Sun, N.~Li, and D.~H.~K. Tsang, ``Optimal {{Scheduling}} of {{Battery Charging Station Serving Electric Vehicles Based}} on {{Battery Swapping}},'' \emph{IEEE Transactions on Smart Grid}, vol.~10, no.~2, pp. 1372--1384, Mar. 2019.

\bibitem{zhu2024}
J.~Zhu, C.~He, K.~Cheung, F.~Luo, and Y.~Liu, ``Low {{Carbon Planning}} of {{Multiple Integrated Energy Systems Considering Trans-Regional Battery Logistics Network}},'' \emph{IEEE Transactions on Sustainable Energy}, vol.~15, no.~2, pp. 1239--1255, Apr. 2024.

\bibitem{qi2023}
W.~Qi, Y.~Zhang, and N.~Zhang, ``Scaling {{Up Electric-Vehicle Battery Swapping Services}} in {{Cities}}: {{A Joint Location}} and {{Repairable-Inventory Model}},'' \emph{Management Science}, vol.~69, no.~11, pp. 6855--6875, Nov. 2023.

\bibitem{mak2013}
H.-Y. Mak, Y.~Rong, and Z.-J.~M. Shen, ``Infrastructure {{Planning}} for {{Electric Vehicles}} with {{Battery Swapping}},'' \emph{Management Science}, vol.~59, no.~7, pp. 1557--1575, Jul. 2013.

\bibitem{sarker2015}
M.~R. Sarker, H.~Pand{\v z}i{\'c}, and M.~A. {Ortega-Vazquez}, ``Optimal {{Operation}} and {{Services Scheduling}} for an {{Electric Vehicle Battery Swapping Station}},'' \emph{IEEE Transactions on Power Systems}, vol.~30, no.~2, pp. 901--910, Mar. 2015.

\bibitem{siddiqua2023}
A.~Siddiqua, V.~Cherala, and P.~K. Yemula, ``Optimal {{Sizing}} and {{Adaptive Charging Strategy}} for the {{Battery Swapping Station}},'' in \emph{2023 {{IEEE PES}} 15th {{Asia-Pacific Power}} and {{Energy Engineering Conference}} ({{APPEEC}})}, Dec. 2023, pp. 1--6.

\bibitem{wang2021}
X.~Wang, J.~Wang, and J.~Liu, ``Vehicle to {{Grid Frequency Regulation Capacity Optimal Scheduling}} for {{Battery Swapping Station Using Deep Q-Network}},'' \emph{IEEE Transactions on Industrial Informatics}, vol.~17, no.~2, pp. 1342--1351, Feb. 2021.

\bibitem{zheng2014}
Y.~Zheng, Z.~Y. Dong, Y.~Xu, K.~Meng, J.~H. Zhao, and J.~Qiu, ``Electric {{Vehicle Battery Charging}}/{{Swap Stations}} in {{Distribution Systems}}: {{Comparison Study}} and {{Optimal Planning}},'' \emph{IEEE Transactions on Power Systems}, vol.~29, no.~1, pp. 221--229, Jan. 2014.

\bibitem{el-taweel2023}
N.~A. {El-Taweel}, A.~Ayad, H.~E.~Z. Farag, and M.~Mohamed, ``Optimal {{Energy Management}} for {{Battery Swapping Based Electric Bus Fleets With Consideration}} of {{Grid Ancillary Services Provision}},'' \emph{IEEE Transactions on Sustainable Energy}, vol.~14, no.~2, pp. 1024--1036, Apr. 2023.

\bibitem{yang2015}
J.~Yang and H.~Sun, ``Battery swap station location-routing problem with capacitated electric vehicles,'' \emph{Computers \& Operations Research}, vol.~55, pp. 217--232, Mar. 2015.

\bibitem{raeesi2020}
R.~Raeesi and K.~G. Zografos, ``The electric vehicle routing problem with time windows and synchronised mobile battery swapping,'' \emph{Transportation Research Part B: Methodological}, vol. 140, pp. 101--129, Oct. 2020.

\bibitem{hof2017}
J.~Hof, M.~Schneider, and D.~Goeke, ``Solving the battery swap station location-routing problem with capacitated electric vehicles using an {{AVNS}} algorithm for vehicle-routing problems with intermediate stops,'' \emph{Transportation Research Part B: Methodological}, vol.~97, pp. 102--112, Mar. 2017.

\bibitem{liang2023}
Y.~Liang, Z.~Ding, T.~Zhao, and W.-J. Lee, ``Real-{{Time Operation Management}} for {{Battery Swapping-Charging System}} via {{Multi-Agent Deep Reinforcement Learning}},'' \emph{IEEE Transactions on Smart Grid}, vol.~14, no.~1, pp. 559--571, Jan. 2023.

\bibitem{mao2024}
S.~Mao, J.~Jin, and Y.~Xu, ``Routing and {{Charging Scheduling}} for {{EV Battery Swapping Systems}}: {{Hypergraph-Based Heterogeneous Multiagent Deep Reinforcement Learning}},'' \emph{IEEE Transactions on Smart Grid}, vol.~15, no.~5, pp. 4903--4916, Sep. 2024.

\bibitem{wan2023}
Y.~Wan, J.~Qin, Y.~Shi, W.~Fu, and D.~Zhang, ``Privacy-{{Preserving Operation Management}} of {{Battery Swapping}} and {{Charging System With Dual-Based Benders Decomposition}},'' \emph{IEEE Transactions on Smart Grid}, vol.~14, no.~5, pp. 3899--3912, Sep. 2023.

\bibitem{rtw_ACC}
R.~Wang, Y.~Ju, Z.~Allybokus, W.~Zeng, N.~Obrecht, and S.~Moura, ``Optimal {{Sizing}}, {{Operation}}, and {{Efficiency Evaluation}} of {{Battery Swapping Stations}} for {{Electric Heavy-Duty Trucks}},'' in \emph{2024 {{American Control Conference}} ({{ACC}})}, Jul. 2024, pp. 707--712.

\bibitem{ample_bss}
\BIBentryALTinterwordspacing
Ample - electric cars for everyone. [Online]. Available: \url{https://ample.com/}
\BIBentrySTDinterwordspacing

\bibitem{saxena2015}
S.~Saxena, C.~Le~Floch, J.~MacDonald, and S.~Moura, ``Quantifying {{EV}} battery end-of-life through analysis of travel needs with vehicle powertrain models,'' \emph{Journal of Power Sources}, vol. 282, pp. 265--276, May 2015.

\bibitem{ansean2013}
D.~Anse{\'a}n, M.~Gonz{\'a}lez, J.~C. Viera, V.~M. Garc{\'i}a, C.~Blanco, and M.~Valledor, ``Fast charging technique for high power lithium iron phosphate batteries: {{A}} cycle life analysis,'' \emph{Journal of Power Sources}, vol. 239, pp. 9--15, Oct. 2013.

\bibitem{NHTS}
\BIBentryALTinterwordspacing
{Federal Highway Administration}, ``{2022 NextGen NHTS National Passenger OD Data U.S. Department of Transportation, Washington, DC.}'' 2022. [Online]. Available: \url{https://nhts.ornl.gov/od/}
\BIBentrySTDinterwordspacing

\bibitem{MSA_area}
\BIBentryALTinterwordspacing
{Employment Development Department, California}, ``Metropolitan {{Areas}} in {{California}}.'' [Online]. Available: \url{https://labormarketinfo.edd.ca.gov/definitions/metropolitan-statistical-areas.html}
\BIBentrySTDinterwordspacing

\bibitem{HEREDeveloper}
\BIBentryALTinterwordspacing
{HERE Developer}, ``{HERE Maps API and SDK Platform Access},'' 2022. [Online]. Available: \url{https://developer.here.com/}
\BIBentrySTDinterwordspacing

\bibitem{basso2019}
R.~Basso, B.~Kulcs{\'a}r, B.~Egardt, P.~Lindroth, and I.~{Sanchez-Diaz}, ``Energy consumption estimation integrated into the {{Electric Vehicle Routing Problem}},'' \emph{Transportation Research Part D: Transport and Environment}, vol.~69, pp. 141--167, Apr. 2019.

\bibitem{weight2022a}
\BIBentryALTinterwordspacing
Massive {EV} batteries deliver big range—and tons and tons of curb weight. Section: News. [Online]. Available: \url{https://www.autoweek.com/news/a39449944/problem-with-ev-battery-weight/}
\BIBentrySTDinterwordspacing

\bibitem{ca_hos}
\BIBentryALTinterwordspacing
``Federal \& {{State Regulations}} {\textbar} {{Hours}} of service - {{California}} {\textbar} {{J}}. {{J}}. {{Keller}}{\textregistered} {{Compliance Network}}.'' [Online]. Available: \url{https://jjkellercompliancenetwork.com/regsense/hours-of-service-california}
\BIBentrySTDinterwordspacing

\bibitem{driving_hour}
\BIBentryALTinterwordspacing
Summary of hours of service regulations {\textbar} {FMCSA}. [Online]. Available: \url{https://www.fmcsa.dot.gov/regulations/hours-service/summary-hours-service-regulations}
\BIBentrySTDinterwordspacing

\bibitem{shi2022}
J.~Shi, M.~Tian, S.~Han, T.-Y. Wu, and Y.~Tang, ``Electric vehicle battery remaining charging time estimation considering charging accuracy and charging profile prediction,'' \emph{Journal of Energy Storage}, vol.~49, p. 104132, May 2022.

\bibitem{speth2022}
D.~Speth, P.~Pl{\"o}tz, S.~Funke, and E.~Vallarella, ``Public fast charging infrastructure for battery electric trucks---a model-based network for {{Germany}},'' \emph{Environmental Research: Infrastructure and Sustainability}, vol.~2, no.~2, p. 025004, Jun. 2022.

\bibitem{caiso}
``{California ISO},'' https://www.caiso.com/.

\bibitem{oasis}
``Open {{Access Same-Time Information System}} {{OASIS - California ISO}},'' {https://www.caiso.com/systems-applications/portals-applications/open-access-same-time-information-system-oasis}, 2024.

\bibitem{watttime}
``{{WattTime}}: {{Data Signals Overview}},'' https://watttime.org/data-science/data-signals/.

\bibitem{nordhaus2017}
W.~D. Nordhaus, ``Revisiting the social cost of carbon,'' \emph{Proceedings of the National Academy of Sciences}, vol. 114, no.~7, pp. 1518--1523, Feb. 2017.

\bibitem{auffhammer2018}
M.~Auffhammer, ``Quantifying {{Economic Damages}} from {{Climate Change}},'' \emph{Journal of Economic Perspectives}, vol.~32, no.~4, pp. 33--52, Nov. 2018.

\bibitem{bressler2021}
R.~D. Bressler, ``The mortality cost of carbon,'' \emph{Nature Communications}, vol.~12, no.~1, p. 4467, Jul. 2021.

\bibitem{gurobi}
\BIBentryALTinterwordspacing
{Gurobi Optimization, LLC}, ``{Gurobi Optimizer Reference Manual},'' 2023. [Online]. Available: \url{https://www.gurobi.com}
\BIBentrySTDinterwordspacing

\bibitem{richard2018a}
L.~Richard and M.~Petit, ``Fast charging station with battery storage system for {{EV}}: {{Grid}} services and battery degradation,'' in \emph{2018 {{IEEE International Energy Conference}} ({{ENERGYCON}})}, 2018, pp. 1--6.

\bibitem{2022bloomberg}
\BIBentryALTinterwordspacing
Lithium-ion battery pack prices rise for first time to an average of \$151/{kWh}. [Online]. Available: \url{https://about.bnef.com/blog/lithium-ion-battery-pack-prices-rise-for-first-time-to-an-average-of-151-kwh/}
\BIBentrySTDinterwordspacing

\bibitem{driver_wage}
{Bureau of Labor Statistics}, ``Occupational employment and wage statistics - heavy and {{Tractor-Trailer Truck Drivers}}.''

\bibitem{cunanan2021}
C.~Cunanan, M.-K. Tran, Y.~Lee, S.~Kwok, V.~Leung, and M.~Fowler, ``A {{Review}} of {{Heavy-Duty Vehicle Powertrain Technologies}}: {{Diesel Engine Vehicles}}, {{Battery Electric Vehicles}}, and {{Hydrogen Fuel Cell Electric Vehicles}},'' \emph{Clean Technologies}, vol.~3, no.~2, pp. 474--489, Jun. 2021.

\bibitem{moultak2017}
M.~Moultak, N.~Lutsey, and D.~Hall, ``Transitioning to zero-emission heavy-duty freight vehicles,'' Tech. Rep., 2017.

\bibitem{michaelnicholas2019}
{Michael Nicholas}, ``Estimating electric vehicle charging infrastructure costs across major {{U}}.{{S}}. metropolitan areas,'' The International Council on Clean Transportation (ICCT), Tech. Rep., 2019.

\bibitem{gamage2023}
T.~Gamage, G.~Tal, and A.~T. Jenn, ``The costs and challenges of installing corridor {{DC Fast Chargers}} in {{California}},'' \emph{Case Studies on Transport Policy}, vol.~11, p. 100969, Mar. 2023.

\end{thebibliography}

\end{document}